\documentclass[twocolumn]{autart}

\usepackage{amsmath} 
\usepackage{amssymb} 
\usepackage{mathrsfs}
\usepackage{empheq}
\usepackage{siunitx}
\usepackage{enumitem}

\newtheorem{theorem}{Theorem}
\newtheorem{problem}{Problem}

\newtheorem{proposition}{Proposition}

\newtheorem{assumption}{Assumption}
\newtheorem{lemma}{Lemma}

\newtheorem{remark}{Remark}
\newtheorem{exx}{Example}
\newenvironment{myproof}{{\it Proof.~}}{\hfill$\diamondsuit$\\\vspace{-0.3cm}}

\newcommand{\vect}[1]{\mathbf{#1}}
\newcommand{\mat}[1]{\mathbf{#1}}
\newcommand{\bs}[1]{\pmb{#1}}
\newcommand{\eps}{\bs{\epsilon}}

\newcommand{\colvec}[2][.95]{%
  \scalebox{#1}{%
    \renewcommand{\arraystretch}{1}%
    $\begin{bmatrix}#2\end{bmatrix}$%
  }
}

\usepackage{color}
\usepackage{float}
\usepackage{subfig}
\usepackage{graphicx}
\usepackage{caption} 
\graphicspath{{images/}}
\definecolor{gray}{rgb}{0.5,0.5,0.5}

\usepackage{hyperref} 
\hypersetup{
    colorlinks,
    citecolor=black,
    filecolor=black,
    linkcolor=black,
    urlcolor=black,
    pdfauthor={},
    pdfsubject={},
    pdftitle={}
}

\usepackage{soul}

\usepackage{textcase}

\begin{document}

\begin{frontmatter}

\title{Hierarchical non-linear control for multi-rotor asymptotic stabilization based on zero-moment direction}

\thanks{This work has been partially funded by: the European Union's Horizon 2020 research and innovation program under grant agreement No 644271 AEROARMS; by the LAAS-CNRS under the grant GRASP and Carnot project; by the University of Padova under grant agreement BIRD168152.}

\author[unipd]{Giulia Michieletto}\ead{giulia.michieletto@unipd.it},
\author[unipd]{Angelo Cenedese}\ead{angelo.cenedese@unipd.it},
\author[laas,unitn]{Luca Zaccarian}\ead{luca.zaccarian@laas.fr}, 
\author[laas]{Antonio Franchi}\ead{antonio.franchi@laas.fr}

\address[unipd]{Department of Information Engineering, University of Padova, Padova, Italy}
\address[laas]{LAAS-CNRS, Universit\'e de Toulouse, CNRS, Toulouse, France}
\address[unitn]{Department of Industrial Engineering, University of Trento, Trento, Italy}

\begin{abstract}
We consider the hovering control problem for a class of multi-rotor aerial platforms with generically oriented propellers.
Given the intrinsically coupled translational and rotational dynamics of such vehicles, we first discuss some assumptions for the considered systems 
to reject torque disturbances and to balance the gravity force, which are translated into a geometric characterization of the platforms that is usually fulfilled by both standard 
models and more general configurations.
%
Hence, we propose a control strategy based on the identification of a zero-moment direction for the applied force and the dynamic state feedback linearization around this preferential direction, which allows to asymptotically stabilize the platform to a static hovering condition.
Stability and convergence properties of the control law are rigorously proved through Lyapunov-based methods and reduction theorems for the stability of nested sets.
%
Asymptotic zeroing of the error dynamics and convergence to the static hovering condition are then confirmed by simulation results on a star-shaped hexarotor model with tilted propellers.
%
%
%
\end{abstract}

\end{frontmatter}

\section{Introduction}\label{sec:intro}

In the last years, technological advances in miniaturized 
sensors/actuators 
and optimized data processing have lead to extensive use of small autonomous flying vehicles within the 
academic, military, and (more recently) commercial contexts (see~\cite{Fuhrmann2017,Shakhatreh2018,Tang2018} and references therein).
Thanks to their high maneuverability and  versatility, Unmanned Aerial Vehicles (UAVs) are rapidly increasing in popularity, thus becoming a mature technology in several application fields ranging from the classical visual sensing tasks (e.g., surveillance and aerial photography~\cite{Kim2018,Motlagh2017} to the recent environment exploration and physical interaction (e.g., search and rescue operations, grasping and manipulation~\cite{Hayat2017,Loianno2018,Ollero2018,Ruggiero2018,Staub2018}). 

In most of these frameworks, the vehicle is required to stably hover in a fixed position. Therefore, many control strategies are known in the literature to enhance the stability of a UAV able to solve this task. These are generally linear solutions based on proportional-derivative schemes or linear quadratic regulators, see, e.g., \cite{Alkhoori2017,Liu2016,Sandiwan2017}.
Hovering non-linear controllers are instead not equally popular and mainly exploit 
feedback 
linearization~\cite{Antonello2018,Lotufo2016},  sliding mode and backstepping techniques~\cite{Abci2017,Chen2016} and/or  geometric control approaches~\cite{Franchi2018,Invernizzi2017a}.

Although less diffused, the effectiveness of the non-linear hovering control schemes has been widely confirmed by experimental tests. For example, in~\cite{Carrillo2012} the performance of controllers based on nested saturations, backstepping and sliding modes has been experimentally evaluated with the aim of stabilizing the position of a quadrotor w.r.t. a visual landmark on the ground. In~\cite{Choi2015}
a quadrotor platform has been used  to validate the possibility of  stably tracking a point through a non-linear control strategy that exploits a backstepping-like feedback linearization method. In~\cite{Goodarzi2017} the experimental results confirm the performance 
of a geometric nonlinear controller during the
autonomous tracking of a Lissajous curve by means of a small quadrotor.

A deep overview of feedback control laws for under actuated UAVs is given in~\cite{Hua2013}, where the authors claim that the non-linear approach to control problems can always be seen as an extension of locally approximated linear solutions. Hence one could derive provable convergence properties by stating some suitable assumptions. 
In this sense, Lyapunov theory has been exploited in~\cite{Lee2010} to prove the convergence of the proposed (non-linear) tracking controller 
assuming bounded initial errors. In detail, the control solution introduced in~\cite{Lee2010} exploits a geometric approach on the three-dimensional Special Euclidean manifold and
ensures the almost global exponential convergence of the tracking error towards the zero equilibrium. A Lyapunov-based approach  is used also in~\cite{Franchi2018} for the more general class of laterally-bounded force aerial vehicles, which includes both under actuated and fully actuated systems with saturations.

In this context, the contribution of our work can be summarized as follows. 
First, we account for a class of multi-rotor aerial platforms having more complex dynamics than the standard quadrotors. More specifically, we address the case where the propellers are in any number (possibly larger than four) and their spinning axes are generically oriented (including the non-parallel case). This  entails the fact that the direction along which the control force is exerted is not necessarily orthogonal to the plane containing all the propellers centers\footnote{{This is strictly valid for standard star-shaped or H-shaped configurations, while for the Y-shaped case and other ones this idea can be easily generalized.}} 
%
and that the control moment is not independent of the control force, as in the typical frameworks
, see, e.g.,~\cite{Lee2010}. For such generic platforms, we propose a non-linear hovering control law that rests upon the identification of a so-called \textit{zero-moment direction}. This concept, introduced in~\cite{Michieletto2017a,Michieletto2017b}, refers to a virtual direction 
along which the intensity of the control force can be freely assigned being the control moment equal to zero.   
The designed controller exploits a sort of dynamic feedback linearization around this preferential direction which is assumed to be generically oriented (contrarily to the state-of-the-art multi-rotor controllers). Its implementation  asymptotically stabilizes the platform to a given constant reference position, constraining its linear and angular velocities to be zero (\textit{static hover condition}~\cite{Michieletto2017b}). The proposed control strategy requires some  algebraic prerequisites on the control matrices that map the motors input to the vehicle control force and torque.  These are fulfilled by the majority of quadrotor models and result to be  non-restrictive so that the designed controller can be applied to both standard 
multi-rotor platforms, whose propellers spinning axes are all parallel, and more general ones.
The convergence properties of the control law are confirmed by the numerical simulations and  are rigorously proved through a Lyapunov-based proof and suitable reduction theorems for the stability of nested sets, extending the results provided in~\cite{Michieletto2017c}. 
%


The rest of the paper is organized as follows. Since we use the unit quaternion representation of the attitude, in
Section~\ref{sec:preliminaries} some basic notions on the related mathematics are given.
In Section~\ref{sec:model} the dynamic model of a generic multi-rotor platform is derived exploiting the Newton-Euler approach. In Section~\ref{sec:control}
the main contribution is provided, presenting the non-linear controller and proving its convergence properties. The theoretical observations are validated
by means of numerical results in Section~\ref{sec:simulation}. Finally, in
Section~\ref{sec:conclusions} some conclusions are drawn and future research directions are discussed. 

\section{Preliminaries and Notation}
\label{sec:preliminaries}

In this work, the unit quaternion formalism is adopted to represent the UAV attitude, overcoming the singularities that characterize Euler angles and simplifying the equations w.r.t. the rotation matrices representation.
 To provide a mathematical background for the model and the controller described hereafter, the main properties of the unit quaternions are recalled in this section.  The reader is referred to~\cite{Diebel2006} and~\cite{Kuipers1999} for further details.

A unit quaternion $\vect{q}$ is a hyper-complex number belonging to the unit 
hypersphere $\mathbb{S}^3$  embedded in $\mathbb{R}^4$. This is usually represented as a four dimensional vector having unitary norm made up of a scalar part, $\eta \in \mathbb{R}$, and a vector part, $\eps \in \mathbb{R}^3$, so that $\vect{q} := \colvec{ \eta & \eps^\top}^\top$ with $ \Vert \vect{q} \Vert^2 = \eta^2 + \Vert \eps \Vert ^2 = 1$.
%
Each unit quaternion $\vect{q}$ corresponds to a unique rotation matrix belonging to the Special Orthogonal group $SO(3) := \{ \mat{R} \in \mathbb{R}^{3 \times 3} \; | \; \mat{R}^\top \mat{R}= \mat{I}_3, \; \text{det}(\mat{R}) = 1\}$. Formally, this is
\begin{align}
\label{eq:q2R}
\mat{R}(\vect{q}) & = \mat{I}_3+2\eta [\eps ]_\times+2[ \eps ] _\times^2  \nonumber\\
&= \mat{I}_3+2\eta [\eps ]_\times + 2 (\eps \eps^\top-\eps^\top \eps \mat{I}_3),
\end{align}
where the operator $[ \cdot]_\times$ denotes the map that associates any non-zero vector in $\mathbb{R}^3$ to the related skew-symmetric matrix in the special orthogonal Lie algebra $\mathfrak{so}(3)$. Thanks to~\eqref{eq:q2R}, it can be verified that $\mat{R}(\vect{q})^\top\mat{R}(\vect{q}) = \mat{R}(\vect{q}_I) = \mat{I}_3$ where $\vect{q}_I:= \colvec{1 & 0 & 0 & 0 }^\top$ is the \textit{identity} (unit) quaternion. 

The claimed relationship is not bijective as each rotation matrix corresponds to two unit quaternions. 
To explain this fact, it is convenient to consider the following axis-angle representation for a unit quaternion, namely $\vect{q} =  \colvec{\cos\left(\frac{\theta}{2} \right) &  \sin\left(\frac{\theta}{2} \right)\!\vect{u}^\top}^\top,$
where $\vect{u} \in \mathbb{S}^2$ identifies the rotation axis and $\theta \in (-\pi, +\pi]$ is the corresponding rotation angle. Using this expression, it can be verified that a rotation around $-\vect{u}$ of an angle $-\theta$ is described by another unit quaternion associated with a rotation by $\theta$ about $\vect{u}$. This feature of the unit quaternions is often referred in literature as \textit{double coverage property}.

In quaternion-based algebra, the rotations composition is performed through the \textit{quaternions product}, denoted hereafter by the symbol $\otimes$. Specifically, given $\vect{q}_1, \vect{q}_2$, it holds that $ \mat{R} (\vect{q}_1) \mat{R} (\vect{q}_2)=\mat{R} (\vect{q}_{3})$, where 
\begin{align}
\label{eq:otimes}
&\vect{q}_3 := \vect{q}_1\otimes\vect{q}_2 = \mat{A}(\vect{q}_1)\vect{q}_2= \mat{B}(\vect{q}_2)\vect{q}_1, 
\end{align}
with
\begin{align}
&\mat{A}(\vect{q}) := \colvec{ \eta & -\eps^\top \\ \eps & \eta\mat{I}_3+[\eps ]_\times }, \quad 
 \mat{B}(\vect{q}):= \colvec{ \eta & -\eps^\top \\ \eps & \eta\mat{I}_3-[\eps ]_\times }.
\end{align}
%
According to~\eqref{eq:otimes}, the inverse of a quaternion $\vect{q}$ may be chosen as $\vect{q}^{-1} = [\eta \; -\eps^\top]^\top$.

Finally, given two 3D coordinate systems $\mathscr{F}_x$ and $\mathscr{F}_y$ such that the unit quaternion $\vect{q}$ indicates the relative rotation from $\mathscr{F}_x$ to $\mathscr{F}_y$, for any vector $\vect{w}$ expressed in $\mathscr{F}_x$ the corresponding vector $\vect{w}^\prime$ in $\mathscr{F}_y$ is computed as 
\begin{align}
\colvec{
0 \\ \vect{w}^\prime } = \vect{q} \otimes \colvec{
0 \\ \vect{w} } \otimes \vect{q}^{-1}.
\end{align}

The time derivative of a unit quaternion $\vect{q}$ is given by
\begin{align}
\label{eq:quat_derivative_B}
\dot{\vect{q}}= \frac{1}{2} \vect{q} \otimes \colvec{
0 \\ \bs{\omega} }  = \frac{1}{2} \mat{A}(\vect{q}) \colvec{
0 \\ \bs{\omega} }   =\frac{1}{2} \colvec{
-\eps^\top \\ \eta \mat{I}_3+[\eps ]_\times
} \bs{\omega},
\end{align}
denoting by $\bs{\omega} \in \mathbb{R}^3$ the angular velocity of $\mathscr{F}_x$ w.r.t. $\mathscr{F}_y$ expressed in $\mathscr{F}_x$.
Relation~\eqref{eq:quat_derivative_B} should be replaced by
\begin{align}
\label{eq:quat_derivative_W}
\dot{\vect{q}} = \frac{1}{2} \colvec{
0 \\ \bs{\omega}^\prime } \otimes \vect{q} = \frac{1}{2} \mat{B}(\vect{q}) \colvec{
0 \\ \bs{\omega}^\prime }   = \frac{1}{2} \colvec{
-\eps^\top \\ \eta \mat{I}_3-[\eps ]_\times
} \bs{\omega}^\prime,
\end{align}
when the angular velocity is expressed in $\mathscr{F}_y$, namely $\bs{\omega}^\prime = \mat{R}(\vect{q})\bs{\omega}$. 

\section{Multi-Rotor Vehicle Dynamic Model}\label{sec:model}

\begin{figure*}[t!]
\begin{center}
\includegraphics[width=0.95\textwidth]{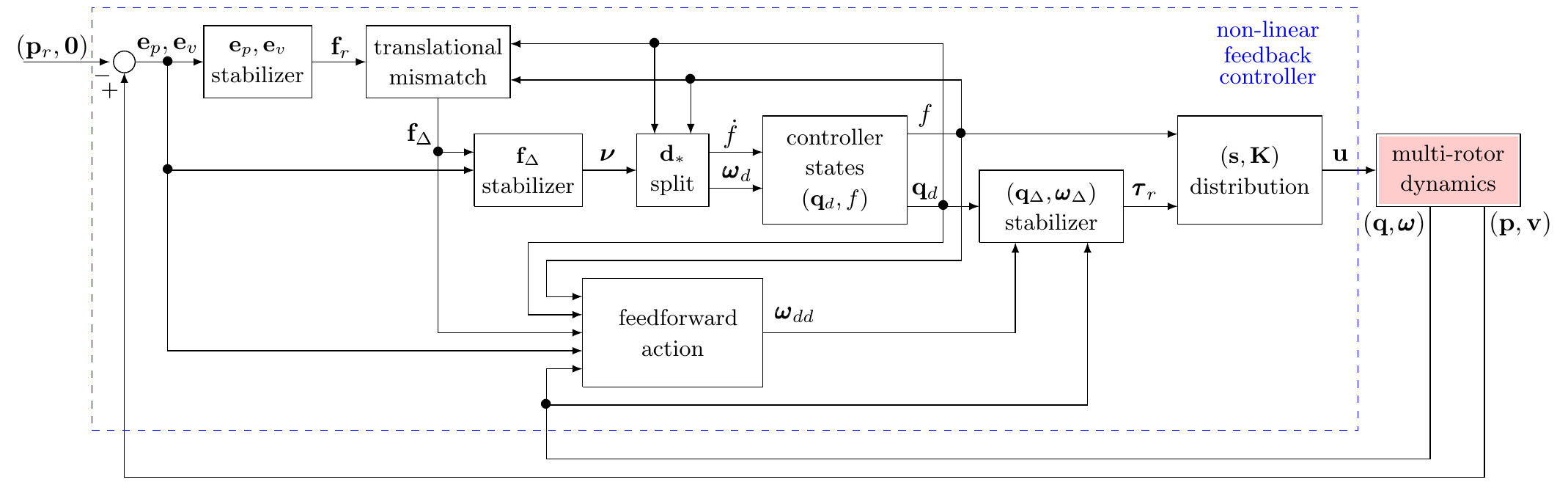}
\caption{Block diagram of the closed-loop system with the proposed dynamic control strategy.}
\label{fig:controller}
\end{center}
\end{figure*}

Consider a generic aerial multi-rotor platform, composed by a rigid body and $n \geq 4$ propellers (with negligible mass and moment of inertia w.r.t. body inertial parameters), each one spinning about a certain axis which could be  generically oriented. The axes mutual orientation, jointly with the number $n$ of rotors, determines if the UAV is an under actuated or a fully actuated system~\cite{Ryll2016}.
This class of vehicles (also known as \textit{Generically Tilted Multi-Rotors}) has been evaluated for the first time in~\cite{Michieletto2017a}, nonetheless we investigate here the derivation of the dynamic model by exploiting the unit quaternion formalism to represent the attitude of the platform. 


We consider the \textit{body frame} $\mathscr{F}_B$ attached to the UAV so that its origin $O_B$ is coincident with the center of mass (CoM) of the vehicle.  The pose of the platform in the inertial \textit{world frame} $\mathscr{F}_W$ is thus described by the pair $(\vect{p},\vect{q}) \in \mathbb{R}^3 \times \mathbb{S}^3$ where the vector $\vect{p} \in \mathbb{R}^3$ denotes the position of $O_B$ in $\mathscr{F}_W$ and the unit quaternion \mbox{$\vect{q} \in \mathbb{S}^3$} represents the orientation of $\mathscr{F}_B$ w.r.t. $\mathscr{F}_W$ (i.e., it corresponds to the relative rotation from body to world frame, therefore
its inverse provides the world coordinates of a  vector expressed in body frame).
The orientation kinematics of the vehicle is governed by~\eqref{eq:quat_derivative_B},
where $\bs{\omega} \in \mathbb{R}^3$ represents the angular velocity of $\mathscr{F}_B$ w.r.t. $\mathscr{F}_W$, expressed in $\mathscr{F}_B$, whereas the linear velocity of $O_B$ in $\mathscr{F}_W$ is denoted by $\vect{v} = \dot{\vect{p}} \in \mathbb{R}^3$.

The $i$-th propeller, $i=1 \ldots n$, rotates with angular velocity $\bs{\omega}_i \in \mathbb{R}^3$ about its spinning axis which passes through the rotor center $O_{P_i}$. The position $\vect{p}_i \in \mathbb{R}^3$ of $O_{P_i}$ and the direction of $\bs{\omega}_i$ are assumed to be constant in $\mathscr{F}_B$. The propeller angular velocity can thus be expressed as $\bs{\omega}_i := \omega_i \vect{z}_{P_i}$ where $\omega_i \in \mathbb{R}$ indicates the (controllable) rotor spinning rate and $\vect{z}_{P_i} \in \mathbb{S}^2$ is a unit vector parallel to the rotor spinning axis.  
While rotating, each propeller exerts a \textit{thrust/lift  force} $\vect{f}_i \in \mathbb{R}^3$  and a \textit{drag moment} $\bs{\tau}_i \in \mathbb{R}^3$, both oriented along the direction defined by $\vect{z}_{P_i}$ and applied in $O_{P_i}$. According to the most commonly accepted model, these two quantities are related to the rotor rate $\omega_i$ by means of the next relations
\begin{align}
\vect{f}_i & = \sigma c_{f_i} \vert \omega_i\vert \omega_i \vect{z}_{P_i} \quad \text{and} \quad 
\bs{\tau}_i  = - c_{\tau_i}^+ \vert \omega_i\vert \omega_i \vect{z}_{P_i}, \label{eq:tau_i}
\end{align}
where $c_{f_i},c_{\tau_i}^+>0$ and $\sigma \in \{-1,1\}$ are constant parameter depending on the shape of the propeller. The propeller is said of counterclockwise (CCW) type if $\sigma = 1$  and of clockwise (CW) type if $\sigma = -1$. Note that for CCW propellers the thrust has the
same direction as the angular velocity vector, whereas for the CW case it has the opposite direction; the drag moment, instead, is always oppositely oriented w.r.t. $\bs{\omega}_i$. 

Introducing $u_i := \sigma \vert \omega_i \vert \omega_i \in \mathbb{R}$ and $c_{\tau_i} := -\sigma c_{\tau_i}^+ \in \mathbb{R}$, relations~\eqref{eq:tau_i} can be rewritten as
\begin{align}
\vect{f}_i & = c_{f_i} u_i \vect{z}_{P_i} \quad \text{and} \quad 
\bs{\tau}_i  = c_{\tau_i} u_i \vect{z}_{P_i}.
\end{align}
The sum of all the propeller forces coincides with the \textit{control force} $\vect{f}_c \in \mathbb{R}^3$ applied at the platform CoM, while the \textit{control moment} $\bs{\tau}_c \in \mathbb{R}^3$ is the sum of the moment contributions due to both the thrust forces and the drag moments. These can be expressed in $\mathscr{F}_B$ as  
\begin{align}
\vect{f}_{c} &\!=\! \textstyle{\sum\limits_{i=1}^n} \vect{f}_i =\textstyle{\sum\limits_{i=1}^n}  c_{f_i}  \vect{z}_{P_i} u_i, \label{eq:f_c} \\ 
\bs{\tau}_{c} &\!=\!\textstyle{\sum\limits_{i=1}^n} (\vect{p}_i \! \times\! \vect{f}_i+\bs{\tau}_i)  = \textstyle{\sum\limits_{i=1}^n} (c_{f_i}\vect{p}_i \!\times \!\vect{z}_{P_i} + c_{\tau_i} \vect{z}_{P_i}) u_i. \label{eq:tau_c} \hspace{0cm}
\end{align}
Defining the \textit{control input vector} \mbox{$\vect{u} = [ u_1 \;  \ldots \; u_n ]^\top \in \mathbb{R}^n$}, ~\eqref{eq:f_c} and~\eqref{eq:tau_c} can be shortened as
\begin{align}
\vect{f}_{c} = \mat{F}\vect{u}  \quad  \text{and} \quad
\bs{\tau}_{c} =\mat{M}\vect{u},
\end{align} 
where $\mat{F}, \mat{M} \in \mathbb{R}^{3 \times n}$ are the \textit{control force input matrix} and the \textit{control moment input matrix}, respectively.

Using the Newton-Euler approach and neglecting the second order effects (e.g., the propeller gyroscopic effects), the
dynamics of the multi-rotor vehicle is governed by the following system of equations 
\begin{empheq}[left = \empheqlbrace]{align}
\dot{\vect{p}} &= \vect{v} \label{eq:motion1} \\
\dot{\vect{q}} &= \frac{1}{2}  \vect{q} \otimes \colvec{
0 \\ 
\pmb{\omega}}  \label{eq:motion2}\\ 
 m\ddot{\vect{p}} &= -mg \vect{e}_3  
+ \mat{R}(\vect{q})\mat{F}\vect{u} \label{eq:motion3} \\
\mat{J} \dot{\bs{\omega}} &= - \bs{\omega} \times \mat{J} \bs{\omega} + \mat{M}\vect{u} \label{eq:motion4} 
\end{empheq}
where $m>0$ is the platform mass, $g>0$ is the gravitational constant, and $\vect{e}_i$ is the $i$-th canonical unit vector in $\mathbb{R}^3$ with $i\in\{1,2,3\}$. The positive definite constant matrix $\mat{J} \in \mathbb{R}^{3 \times 3}$ describes the vehicle inertia in $\mathscr{F}_B$. 




\section{Zero-moment Force Direction Controller}
\label{sec:control}

In this section we design a non-linear control law to stabilize in static hover conditions an aerial vehicle belonging to the  generic class of multi-rotor platforms described in Section~\ref{sec:model}, namely we solve the following problem. 

\begin{problem}
\label{prob:hovering}
Given plant~\eqref{eq:motion1}-\eqref{eq:motion4},
 find a (possibly dynamic) state feedback control law that assigns the input~$\vect{u}$ to
ensure that, for any constant reference position $\vect{p}_r \! \in \! \mathbb{R}^3$,
the closed-loop system is able to asymptotically stabilize $\vect{p}_r$ with some hovering orientation. In other words, the controller is required to asymptotically stabilize a set where $\vect{p}\!=\! \vect{p}_r$, and $\dot{\vect{p}}$ and $\bs{\omega}$ are both zero, while orientation $\vect{q}$ could be arbitrary but constant.
\end{problem}

The arbitrariness of the orientation is fundamental for the feasibility of Problem~\ref{prob:hovering}, which is in general solvable only if certain steady-state attitudes are realized by the platform (\textit{static hoverability realizability}~\cite{Michieletto2017b}). Nevertheless, a solution can always be found whether matrices $\mat{F}$ and $\mat{M}$ satisfy some suitable properties. For this reason, in Section~\ref{sec:assumptions} some possibly restrictive assumptions (even though some of them can actually be proven to be necessary) are stated. Then in Section~\ref{sec:controller} we illustrate the dynamics and interconnections of the proposed control scheme, represented in Figure~\ref{fig:controller}. The description of this controller is a contribution of our preliminary work~\cite{Michieletto2017c}. Sections~\ref{sec:errordyn}  and~\ref{sec:ext} instead represent the innovative part. We first provide a rigorous proof of asymptotic stability of the error dynamics exploiting a hierarchical structure and the reduction theorems presented in~\cite{Maggiore2013}. Then, we propose an extension of the proposed control law, accounting also for the stabilization of a given constant orientation. 

\subsection{Main Assumption and Induced Zero-moment Direction} 
\label{sec:assumptions}

%

In order to attain constant position and orientation for the platform, the stabilizing controller given in this section requires that the system is able to both reject torque disturbances in any direction and compensate the gravity force. These requirements are satisfied when the next assumption is in place, as proved in the following.
%
%
\begin{assumption} \label{as:standing}
Let $\mat{F}$ and $\mat{M}$ be the control input matrices introduced in~\eqref{eq:f_c}-\eqref{eq:tau_c}, we define matrix $\bar{\mat{F}} $ such that $\mathrm{Im}(\bar{\mat{F}})=\ker(\mat{F})$.
We assume that 
$\emph{rk}(\mat{M}\bar{\mat{F}}) = 3$.
\end{assumption}
%
Assumption~\ref{as:standing} implies $\textrm{rk}(\mat{M}) = 3$, corresponding to the possibility to freely assign the control moment $\bs{\tau}_c$ in a sufficiently large open space of $\mathbb{R}^3$ containing the origin. This is equivalent to requiring full-actuation of the orientation dynamics~\eqref{eq:motion4}, guaranteeing that the platform is able to reject torque disturbances in any direction\footnote{Differently from~\cite{Michieletto2017b}, no constraint is  imposed here on the positivity of the control input vector.}.  

\begin{proposition}
\label{prop:1}
Under Assumption~\ref{as:standing}, the control moment input matrix $\mat{M}$ is full-rank. 
\end{proposition}
%
%
\begin{myproof}
Since $\textrm{rk}(\mat{M}\bar{\mat{F}}) \leq  \min\{\textrm{rk}(\mat{M}),n-\textrm{rk}(\mat{F})\}$ and $\mat{M}$ has three rows, Assumption~\ref{as:standing} yields $\textrm{rk}(\mat{M}) = 3$.
\end{myproof}

Assumption~\ref{as:standing} also entails that $n-\textrm{rk}(\mat{F}) \ge 3$ and $\textrm{rk}([\mat{F}^\top \,|\, \mat{M}^\top]) \geq 4$. 
This results in the existence of at least a unit vector in $\mathbb{R}^n$ (i.e., a direction in the control input space) that generates a zero control moment and, at the same time, identifies a non-zero control force direction. These observations are  formalized in the following proposition and lemma.

\begin{proposition}
Under Assumption~\ref{as:standing},  $\emph{rk}(\mat{F}\bar{\mat{M}}) \geq 1$ for any matrix $\bar{\mat{M}}$ such that $\mathrm{Im}(\bar{\mat{M}})=\ker(\mat{M})$.
\end{proposition}
\begin{myproof}
Ab absurdo, let assume that $\textrm{rk}(\mat{F}\bar{\mat{M}}) = 0$, i.e., the product $\mat{F}\bar{\mat{M}}$ is a null matrix. This implies that $\ker(\mat{M}) \subseteq \ker(\mat{F})$, namely $\ker(\mat{M}) \cap \ker(\mat{F}) = \ker(\mat{M})$.
Recall now that for generic matrices $\mat{A}$ and $\mat{B}$ of suitable dimensions it holds $\mathrm{rk}(\mat{AB}) = \dim(\mathrm{Im}(\mat{AB})) = \textrm{rk}(\mat{B}) - \dim (\ker(\mat{A}) \cap \mathrm{Im}(\mat{B}))$~\cite{Wonham1985}.
Since $\textrm{rk}(\mat{M}) = 3$ from Proposition~\ref{prop:1}, we may write
\begin{align}
\textrm{rk}(\mat{M}\bar{\mat{F}}) 
&= \textrm{rk}(\bar{\mat{F}}) - \textrm{dim} \left( \ker (\mat{M}) \cap \textrm{Im}(\bar{\mat{F}}) \right)\\
&= \textrm{dim} \left( \ker (\mat{F}) \right) - \textrm{dim} \left( \ker (\mat{M}) \right)\\
&= n - \textrm{rk} (\mat{F}) - \left(n - \textrm{rk} (\mat{M}) \right)\\
&= 3 - \textrm{rk}(\mat{F}).
\end{align}
As $\textrm{rk}(\mat{M}\bar{\mat{F}})= 3$, from Assumption~\ref{as:standing}, it should be $\textrm{rk}(\mat{F})=0$ but $\mat{F}$ is nonzero by construction.
\end{myproof}
\begin{lemma}
\label{lemma:equivalence_2}
For the control input matrices $\mat{F}$ and $\mat{M}$ in~\eqref{eq:f_c}-\eqref{eq:tau_c} the following requirements are equivalent:
\begin{enumerate}[leftmargin=0.5cm]
\item[a)] $\emph{rk}(\mat{F}\bar{\mat{M}}) \geq 1$, where $\bar{\mat{M}}$ is such that \mbox{$\mathrm{Im}(\bar{\mat{M}})=\ker(\mat{M})$};
\item[b)] $\exists \bar{\vect{u}} \in \ker(\mat{M})$ such that $\Vert \mat{F} \bar{\vect{u}} \Vert = 1$.
\end{enumerate}
\end{lemma}
\begin{myproof}
\noindent
{\it a) $\Rightarrow$ b)}.  
Since $\text{Im}(\bar{\mat{M}})=\ker(\mat{M})$, one can always  select a unit vector ${\vect{u}}^\star \in \ker(\mat{M})$ as a linear combination of the columns of $\bar{\mat{M}}$ and the rank condition ensures that $\mat{F} {\vect{u}}^\star \neq \vect{0}$. Choosing $\bar{\vect{u}} = \vect{u}^\star /\Vert \mat{F} \vect{u}^\star \Vert$ completes the proof.
\smallskip\\
\noindent
{\it b) $\Rightarrow$ a)}. The existence of $\bar{\vect{u}} \in \ker(\mat{M})$ implies that $\ker(\mat{M}) = \text{Im}(\bar{\mat{M}})\neq \emptyset$. Moreover,  from $\mat{F} \bar{\vect{u}} \neq \vect{0}$, it is guaranteed that \mbox{$\textrm{rk}(\mat{F}\bar{\mat{M}}) \geq 1$}.
\end{myproof}

The starting point of the proposed controller is the identification of a direction in the force space  along which the intensity $\Vert \vect{f}_c \Vert$ of the control force can be arbitrarily assigned when the control moment $\bs{\tau}_c$ is equal to zero. This \textit{zero-moment preferential direction}, identified by $\vect{d}_* \in \text{Im}(\mat{F})\cap \mathbb{S}^2$, has thus to be defined based on the null space of $\mat{M}$. Using Assumption~\ref{as:standing} and its implications in Lemma~1, a suitable choice is 
\begin{align}
\label{eq:d_*}
\vect{d}_*= \mat{F}\bar{\vect{u}}.
\end{align}

%
Finally, we can observe that Assumption~\ref{as:standing} entails that the product $\mat{M}\bar{\mat{F}}$ is right-invertible, namely there exists a matrix $\mat{X}$, whose dimensions depends on the rank of $\mat{F}$, such that  $\mat{M}\bar{\mat{F}} \mat{X} = \mat{I}_3$.  This constraint 
is equivalent to the property introduced in our preliminary work~\cite{Michieletto2017c} implying the existence of a generalized right pseudo-inverse of $\mat{M}$ as formally stated in the next lemma. 

\begin{lemma}
\label{lemma:equivalence}
Assumption~\ref{as:standing} holds if and only if   $\exists \mat{K} \in \mathbb{R}^{n \times n}$ such that
$\mat{M} \mat{K} \mat{M}^\top$ is invertible and 
 $\mat{F} \mat{M}^\dagger_K = \mat{0}$, where $\mat{M}^{\dagger}_K=\mat{K}\mat{M}^\top   (\mat{M} \mat{K} \mat{M}^\top)^{-1}  \in \mathbb{R}^{n \times 3}$ is the generalized right pseudo-inverse of $\mat{M}$.
\end{lemma}
\begin{myproof} 
\noindent
{\it$\Rightarrow$} Assume $\textrm{rk}(\mat{M}\bar{\mat{F}}) =3$. Then, selecting
$\mat{K}:=\bar{\mat{F}}(\bar{\mat{F}})^\top$ we obtain from the rank condition 
that $\mat{M} \mat{K} \mat{M}^\top = \mat{M} \bar{\mat{F}} (\mat{M} \bar{\mat{F}})^\top \in {\mathbb R}^{3\times 3}$ is invertible. Moreover $\mat{F} \mat{M}^\dagger_K = \mat{0}$ because $\mat{F} \bar{\mat{F}} = \mat{0}$.
\smallskip\\
\noindent
{\it$\Leftarrow$} Proceeding ab absurdo, let us assume $rk(\mat{M}\bar{\mat{F}}) < 3$ and that a matrix $\mat{K}$ exists satisfying the properties in the statement of the lemma; for that matrix we have 
\begin{equation}
\label{eq:goodKabsurd}
\mat{F} \mat{M}^\dagger_K = \mat{0}, \quad \mat{M} \mat{M}^\dagger_K = \mat{I}.
\end{equation}
Consider now any nonzero $\bs{\tau}_r \notin \text{Im}(\mat{M}\bar{\mat{F}})$ (its existence is guaranteed by the stated rank assumption) and denote $\vect{u} := \mat{M}^\dagger_K \bs{\tau}_r$. Then the left inequality of \eqref{eq:goodKabsurd} implies that $\vect{u} \in \ker (\vect{F})$, i.e., there exists $\vect{w} \in \mathbb{R}^n$ such that $\vect{u} = \bar{\mat{F}} \vect{w}$. Using the right equation in~\eqref{eq:goodKabsurd}, through simple substitutions, we get $ \bs{\tau}_r= \mat{M} \mat{M}^\dagger_K \bs{\tau}_r = \mat{M}\vect{u} = \mat{M}\bar{\mat{F}} \vect{w},$
which clearly contradicts the assumption $\bs{\tau}_r \notin \text{Im}(\mat{M}\bar{\mat{F}})$, leading to an absurd and completing the proof.
\end{myproof}

%
{\begin{remark}
Assumption~\ref{as:standing} essentially enables a sufficient level of decoupling between $\vect{f}_c$ and $\bs{\tau}_c$ ensuring the possibility to identify (at least) a direction along which the control force can be freely assigned guaranteeing zero control moment. Referring to the nomenclature introduce in~\cite{Michieletto2017b}, Assumption~\ref{as:standing} are fulfilled for platforms having at least a decoupled force direction (D1). 
\end{remark}}

\subsection{Controller Scheme}
\label{sec:controller}

Based on Assumption~\ref{as:standing} and its implications in Lemma 2, we propose here a dynamic controller where the control input  $\vect{u}$ is selected as
\begin{align}
\label{eq:u}
\vect{u} = \mat{M}_K^\dagger \bs{\tau}_r + \bar{\vect{u}}  f,
\end{align} 
so that $\bs{\tau}_r \in {\mathbb R}^3$ and $f \in \mathbb{R}$ appear conveniently in the expression of the
control force and the control moment~\mbox{\eqref{eq:f_c}-\eqref{eq:tau_c}} implying, by virtue of Lemma 1 and Lemma 2, 
\begin{align}
\label{eq:explicit_forces_1}
 \vect{f}_c &= \mat{F} \vect{u} = \vect{d}_* f, \\
 \label{eq:explicit_forces_2}
 \bs{\tau}_c &= \mat{M} \vect{u} = \bs{\tau}_r,
\end{align}
which clearly reveals a nice decoupling in the wrench components. 
Once this decoupling is in place, we are interested in steering the platform towards a \textit{desired orientation} $\vect{q}_d$ such that the direction of the resulting force $\mat{R}(\vect{q}_d)\vect{f}_c$ acting on the translational dynamics~\eqref{eq:motion3} (i.e., the direction of $\mat{R}(\vect{q}_d)\vect{d}_*$ because of~\eqref{eq:explicit_forces_1}) coincides with a desired direction arising from a simple PD + gravity compensation feedback function. This is here selected as
\begin{align}
\label{eq:f_r}
\vect{f}_r & := mg\vect{e}_3 -k_{pp} \vect{e}_p -k_{pd} \vect{e}_v ,
\end{align}
where $\vect{e}_p = \vect{p}-\vect{p}_r$ and  $\vect{e}_v =  \vect{v}$ are the \textit{position error} and the \textit{velocity error}, respectively, while $k_{pp},k_{pd} \in \mathbb{R}^+$ are arbitrary (positive) scalar PD gains. Rather than computing $\vect{q}_d$ directly, an auxiliary state can be introduced in the controller,
evolving in ${\mathbb S}^3$ through the  quaternion-based dynamics in~\eqref{eq:quat_derivative_B}, namely
\begin{align}
\label{eq:q_d_dot}
\dot{\vect{q}}_d  = \frac{1}{2} \vect{q}_d \otimes \colvec{ 0 \\ \bs{\omega}_d }, 
\end{align}
where $\bs{\omega}_d \in \mathbb{R}^3$ is an additional virtual input that should be selected so that the actual input to the translational dynamics~\eqref{eq:motion3} eventually converges to the state feedback~\eqref{eq:f_r}. In other words, $\bs{\omega}_d$ should be set
to drive to zero the following mismatch, motivated by~\eqref{eq:motion3} and~\eqref{eq:explicit_forces_1},
\begin{align}
\vect{f}_\Delta :=  \mat{R}(\vect{q}_d) \vect{f}_c -\vect{f}_r =\mat{R}(\vect{q}_d) \vect{d}_*  f -\vect{f}_r.
\label{eq:f_Delta}
\end{align}
%
We will show that such a convergence is ensured by considering the variable $f$ in~\eqref{eq:u} as an additional scalar state of the controller, and then imposing
\begin{align}
\label{eq:omega_d}
\bs{\omega}_d &= \frac{1}{f} \left[ \vect{d}_*\right]_\times \mat{R}^\top(\vect{q}_d) \bs{\nu},\\
\label{eq:f_dot}
\dot{f} & = \left( \mat{R}(\vect{q}_d) \vect{d}_*\right)^\top \bs{\nu},
\end{align}
where 
\begin{align}
\label{eq:nu}
\bs{\nu} \!:= \! 
\left(\frac{k_{pd} k_{pp}}{m}\vect{e}_p+\left(\!  \frac{k_{pd}^2}{m}-k_{pp}\right) \vect{e}_v 
 - \left(\frac{k_{pd} }{m}+k_\Delta \right) \vect{f}_\Delta \right)\! ,
\end{align}
being $k_\Delta \in \mathbb{R}^+$  an additional (positive) scalar gain. Note that equation~\eqref{eq:omega_d} clearly makes sense only if $f \neq 0$ (this is guaranteed by the stated assumptions and  will be formally established in Fact~\ref{fact:A0good} in Section~\ref{sec:stability}).

The scheme is completed by an appropriate selection of $\bs{\tau}_r$ in~\eqref{eq:u} ensuring that the attitude $\vect{q}$ tracks the desired attitude $\vect{q}_d$. This task is easily realizable because of Assumption~\ref{as:standing}, which guarantees the full-authority control action on the rotational dynamics. To simplify the exposition, we introduce the mismatch  $\vect{q}_\Delta \in \mathbb{S}^3$ between the current and the desired orientation, namely 
\begin{align}
\label{eq:q_Delta}
\vect{q}_\Delta:= \vect{q}_d^{-1} \!\otimes\! \vect{q}
&=\colvec{ \eta_d\eta+\eps_d^\top\eps \\ -\eta \eps_d+\eta_d\eps-[\eps_d]_\times \eps } =\colvec{ \eta_\Delta \\ \eps_\Delta }.
\end{align}
Then the \textit{reference moment} $\bs{\tau}_r$ in~\eqref{eq:u} entailing the convergence to zero of this mismatch is given by
\begin{align}
\label{eq:tau_r}
\bs{\tau}_r = - k_{ap} \eps_\Delta  - k_{ad}\bs{\omega}_\Delta+ \bs{\omega} \times \mat{J} \bs{\omega}  + \mat{J} {\bs{\omega}}_{dd},
\end{align} 
where $\bs{\omega}_\Delta=\bs{\omega}-\bs{\omega}_d \in \mathbb{R}^3$ is the angular velocity mismatch and the PD gains $k_{ap}\in \mathbb{R}^+$ and $k_{ad}\in \mathbb{R}^+$ allow tuning the proportional and derivative action of the attitude transient, respectively. 

\begin{figure*}[t!]
\begin{align}
\label{eq:omega_dd}
{\bs{\omega}}_{dd} &= \frac{1}{f}  [\vect{d}_*]_\times  \mat{R}^\top(\vect{q}_d) \left( k_1 \mat{R}(\mat{q})\vect{d}_* \xi f +  k_2(\vect{e}_p,\vect{e}_v,\vect{f}_\Delta)  \vect{e}_p +
k_3(\vect{e}_p,\vect{e}_v,\vect{f}_\Delta)  \vect{e}_v + k_4(\vect{e}_p,\vect{e}_v,\vect{f}_\Delta)  {\vect{f}}_\Delta \right), \quad \text{where} \\
&k_1 =  \frac{k_{pd}^2}{m^2}-\frac{k_{pp}}{m}, \label{eq:k_1}\\
&k_2(\vect{e}_p,\vect{e}_v,\vect{f}_\Delta)  = - \left(  \frac{k_{pd}^2 k_{pp}}{m^2}+\frac{k_{pp}^2}{m} + \kappa(\vect{e}_p,\vect{e}_v,\vect{f}_\Delta)\; \frac{k_{pd} k_{pp}}{m}\right),\\ &k_3(\vect{e}_p,\vect{e}_v,\vect{f}_\Delta)  =  - \left(  \frac{k_{pd}^2 k_{pp}}{m^2}+\frac{k_{pp}^2}{m} + \kappa(\vect{e}_p,\vect{e}_v,\vect{f}_\Delta)\; \frac{k_{pd} k_{pp}}{m}\right), \\
&k_4(\vect{e}_p,\vect{e}_v,\vect{f}_\Delta)   =  \frac{k_{pd}^2}{m^2}-\frac{k_{pp}}{m} +\frac{k_{pd}k_\Delta }{m}+k_\Delta^2 + \kappa(\vect{e}_p,\vect{e}_v,\vect{f}_\Delta)\;\left(\frac{ k_{pd} }{m}+k_\Delta \right),  \label{eq:k_4} \\
\label{eq:kappa}
&\kappa(\vect{e}_p,\vect{e}_v,\vect{f}_\Delta)= -\frac{2}{f}\vect{d}_*^\top  \mat{R}^\top(\vect{q}_d) \left(\frac{k_{pd} k_{pp}}{m}\vect{e}_p+\left(\!  \frac{k_{pd}^2}{m}-k_{pp}\right) \vect{e}_v 
 - \left(\frac{k_{pd} }{m}+k_\Delta \right) \vect{f}_\Delta \right).
\end{align}
\vspace{-0.3cm}
\hrule
\vspace{0.0cm}
\end{figure*}
In~\eqref{eq:tau_r}, a feedforward term clearly appears, compensating for the quadratic terms in $\bs{\omega}$ emerging in~\eqref{eq:motion4}, in addition to a correction term ${\bs{\omega}}_{dd} \in \mathbb{R}^3$ ensuring the forward invariance of the set where 
$\vect{q}=\vect{q}_d$ and $\bs{\omega} = \bs{\omega}_d$. The expression of this term is reported in equation~\eqref{eq:omega_dd} at the top of the next page and can be proved to be equal to $\dot{\bs{\omega}}_d$ along solutions (the proof is
available in the Appendix).

\subsection{Error dynamics}
\label{sec:errordyn}

To analyze the closed-loop system presented in the previous section, the following relevant dynamics are introduced for the \textit{orientation error} variable $\vect{q}_\Delta$ in~\eqref{eq:q_Delta} and the associated \textit{angular velocity mismatch} $\bs{\omega}_\Delta$, i.e., 
\begin{align}
\label{eq:q_Delta_dot}
\dot{\vect{q}}_\Delta &= \frac{1}{2} \vect{q}_\Delta \otimes \colvec{
0 \\ \bs{\omega}_\Delta
}, \\
\label{eq:omega_Delta_dot}
\mat{J} \dot{\bs{\omega}}_\Delta &= -\bs{\omega} \times \mat{J}\bs{\omega}- \mat{J} \dot{\bs{\omega}}_{d} +\bs{\tau}_r.
\end{align}
%
To establish useful properties of the translational dynamics, we evaluate the (translational) error vector \mbox{$\vect{e}_t := 
\colvec{\vect{e}_p^\top & \vect{e}_v^\top 
}^\top \in \mathbb{R}^6$}, 
which well characterizes the deviation from the reference position $\vect{p}_r \in \mathbb{R}^3$. Combining equation~\eqref{eq:motion3} with the definition of $\vect{f}_\Delta$ given in~\eqref{eq:f_Delta} the dynamics of $\vect{e}_t$ can be written as follows
\begin{align}
\label{eq:poserror_dyn1}
\dot{\vect{e}}_p &= \vect{e}_v \\
\label{eq:poserror_dyn2}
m \dot{\vect{e}}_v &= -mg\vect{e}_3 + (\mat{R}(\mat{q}) - \mat{R}(\vect{q}_d)) \vect{f}_c + \vect{f}_r + \vect{f}_\Delta.
\end{align}
%
A last mismatch variable that needs to be characterized is the (scalar) controller state $f$. Combining~\eqref{eq:motion3} with~\eqref{eq:explicit_forces_1}, one realizes that the zero position error condition $\vect{e}_p = \vect{0}$ can only be reached if the state $f$, governed by~\eqref{eq:f_dot}, converges to $mg$. Instead of describing the error system in terms of the deviation $f - mg$ (which should clearly go to zero), we prefer to use the redundant set of coordinates $\vect{f}_\Delta$ in~\eqref{eq:f_Delta}. Indeed, according to~\eqref{eq:f_Delta}, showing that $\vect{f}_\Delta$ tends to zero implies that, asymptotically, we get $\mat{R}(\vect{q}_d) \vect{d}_* f = \vect{f}_r$. Namely, as long as $\vect{e}_t$ tends to zero too, we approach the set where $\vect{d}_* f = mg \mat{R}^\top\!(\vect{q}_d) \vect{e}_3$. 
Note that $\vect{q}_{\Delta}=\vect{q}_I$ implies $\mat{R}(\vect{q}) = \mat{R}(\vect{q}_d)$, this clearly corresponds to the set characterized in Problem~\ref{prob:hovering} where the orientation satisfies $\mat{R}(\vect{q}) \vect{d_*} = \mat{R}(\vect{q}_d) \vect{d_*} =\vect{e}_3$ and $ |f| = mg$.


In the next section we study the stabilizing properties induced by the proposed controller, by relying on the error coordinates introduced above.

\subsection{Stability analysis}
\label{sec:stability}

The error variables, whose closed-loop dynamics has been characterized in the previous section, can be used 
to prove that the proposed control scheme solves Problem~\ref{prob:hovering}. To formalize this observation, let consider the following coordinates for the overall closed loop
\begin{align}
\label{eq:xidef}
\vect{z} &:= (\vect{q}_\Delta,\bs{\omega}_\Delta, \vect{f}_\Delta, \vect{e}_t, \vect{q}) \in \mathcal{Z} \subseteq \mathbb{R}^{20},
\end{align}
and the next compact set (that results from the Cartesian product of compact sets)
\begin{align}
\label{eq:A0}
{\mathcal Z}_0 &:= 
\big\{ \vect{z} \in \mathcal{Z} \;| \; \vect{q}_\Delta = \vect{q}_I, \bs{\omega}_\Delta=\vect{0}, \vect{f}_\Delta=\vect{0},  \nonumber\\
&\hspace{3.5cm} \vect{e}_t=\vect{0}, \mat{R}(\vect{q}) \vect{d_*} =\vect{e}_3 \big\},
\end{align}
which clearly characterizes the requirement that the desired position is asymptotically reached ($\vect{e}_t=\vect{0}$) with some constant orientation, by  
ensuring that the zero-moment direction $\vect{d}_*$ is correctly aligned with the steady-state action $mg\vect{e}_3$, thus compensating the gravity force. 

Before proceeding with the proof, we establish a useful property of the compact set  ${\mathcal Z}_0$ in terms of the fact that the controller state $f$ is non-zero.
\begin{fact} 
\label{fact:A0good}
It exists a neighborhood of the compact set ${\mathcal Z}_0$ where variable $f$ is (uniformly) bounded away from zero.
\end{fact}
\begin{myproof}
Since in ${\mathcal Z}_0$ we have $\vect{e}_t=\vect{0}$ and $\vect{f}_\Delta=\vect{0}$, then 
from \eqref{eq:f_Delta} it follows that
 $\vect{d}_*  f = mg \mat{R}^\top\!(\vect{q}_d) \vect{e}_3$. Taking norm on both sides and due to the property of rotation matrices, it holds that $|f| ={mg}$. Since ${\mathcal Z}_0$ is compact, by continuity there exists a neighborhood of ${\mathcal Z}_0$ where $|f|$ is (uniformly) positively lower bounded.
\end{myproof}

We carry out our stability proof by focusing on increasingly small nested sets, each of them characterized by a desirable behavior of certain components of the variable $\vect{z}$ in~\eqref{eq:xidef}. The first set corresponds to the set where the attitude mismatch $(\vect{q}_\Delta ,\bs{\omega}_\Delta )$ is null. It is defined as
\begin{align}
 {\mathcal Z}_{a} := 
\left\{ \vect{z} \in {\mathcal Z} \; | \; \vect{q}_\Delta = \vect{q}_I, \;  \bs{\omega}_\Delta= \vect{0}    \right\},
\end{align}
and is clearly an unbounded and closed set.
For this set, we may prove 
that solutions remaining close to the compact set ${\mathcal Z}_0$ are well behaved in terms of asymptotic stability of the non-compact set $ {\mathcal Z}_{a}$.
\begin{lemma} \label{lem:A_a}
Set $ {\mathcal Z}_{a}$ is locally asymptotically stable 
near ${\mathcal Z}_0$
for the closed-loop dynamics.
\end{lemma}
\begin{myproof}
We prove the result exploiting the dynamics of variables $\vect{q}_\Delta$ and 
$\bs{\omega}_\Delta$ in \eqref{eq:q_Delta_dot} and \eqref{eq:omega_Delta_dot}. In particular, defining the Lyapunov function 
\begin{align}
V_a &:= 2 k_{ap} (1 -\eta_\Delta) + \frac{1}{2}\bs{\omega}_\Delta^\top \mat{J} \bs{\omega}_\Delta,
\end{align}
which is positive definite in a neighborhood of $ {\mathcal Z}_{a}$.
Using equations~\eqref{eq:tau_r}, \eqref{eq:q_Delta_dot}, \eqref{eq:omega_Delta_dot},
which hold close to ${\mathcal Z}_0$ due to the result established in Fact~\ref{fact:A0good},
 we obtain the dynamics restricted to variables $\vect{q}_\Delta$ and 
$\bs{\omega}_\Delta$, corresponding to
\begin{align}
\dot{\vect{q}}_\Delta &= \colvec{
\dot{\eta}_\Delta \\ \dot{\eps}_\Delta
}= \frac{1}{2}  \vect{q}_\Delta \otimes \colvec{
0 \\ \bs{\omega}_\Delta
}, \\
\mat{J} \dot{\bs{\omega}}_\Delta &= - k_{ap} \eps_\Delta  - k_{ad} \bs{\omega}_\Delta,
\end{align}
which is clearly autonomous (independent of external signals).
Then, the derivative of $V_a$ along the dynamics turns out to be
\begin{align}
\dot{V}_a & = -2k_{ap} \dot{\eta}_\Delta + \bs{\omega}_\Delta^\top \mat{J} \dot{\bs{\omega}}_\Delta\\
& = k_{ap} \bs{\omega}_\Delta^\top \eps_\Delta + \bs{\omega}_\Delta^\top (- k_{ap} \eps_\Delta  - k_{ad} \bs{\omega}_\Delta)\\
&= - k_{ad} \Vert \bs{\omega}_\Delta \Vert ^2 . 
\label{eq:dot_Va}
\end{align}
%
Since the dynamics is autonomous, and the set where both $\vect{q}_\Delta$ and 
$\bs{\omega}_\Delta$ are zero is compact in these restricted coordinates, local asymptotic stability follows from local positive definiteness of $V_a$ and   invariance principle.
\end{myproof}

Establishing asymptotic stability of $ {\mathcal Z}_{a}$ near ${\mathcal Z}_0$, clearly implies its forward invariance near ${\mathcal Z}_0$. Therefore it makes sense to describe the dynamics of the closed loop restricted to this set, which is easily computed by replacing $\vect{q}_d$ with $\vect{q}$ and $\bs{\omega}_d$ by $\bs{\omega}$ wherever they appear.

The next step is then to prove asymptotic stability of 
\begin{align}
{\mathcal Z}_f := 
\left\{ \vect{z} \in  {\mathcal Z}_{a} \; | \; \vect{f}_\Delta = 0    \right\},
\end{align}
i.e., the set where the virtual input $\vect{f}_r$ in~\eqref{eq:f_r} is the actual input of the translational dynamics~\eqref{eq:motion1}.
%
Its asymptotic stability near ${\mathcal Z}_0$ is established next for initial conditions in $ {\mathcal Z}_{a}$.

\begin{lemma}
 \label{lem:A_f}
Set ${\mathcal Z}_f$ is asymptotically stable near ${\mathcal Z}_0$ for the closed-loop dynamics with initial conditions in $ {\mathcal Z}_{a}$.
\end{lemma}
\begin{myproof}
Consider the derivative of variable $\vect{f}_\Delta$, along dynamics \eqref{eq:poserror_dyn1}-\eqref{eq:poserror_dyn2} restricted to $ {\mathcal Z}_{a}$ (namely such that $\vect{q}=\vect{q}_d$). Using the definition in \eqref{eq:f_Delta}, we obtain 
\begin{align}
\dot{\vect{f}}_\Delta & = \mat{R}(\vect{q}_d) \vect{d}_*  \dot{ f}+\dot{\mat{R}}(\vect{q}_d) \vect{d}_*  f  - \dot{\vect{f}}_r \\
&= \dot{\vect{f}}_{\Delta,1} +\dot{\vect{f}}_{\Delta,2} +\dot{\vect{f}}_{\Delta,3} 
\end{align}
\begin{align} 
\dot{\vect{f}}_{\Delta,1} & =\mat{R}(\vect{q}_d) \vect{d}_*  \dot{ f}  = \left(\mat{R}(\vect{q}_d) \vect{d}_*\right)  \left( \mat{R}(\vect{q}_d) \vect{d}_*\right)^\top \bs{\nu}  \\
& = \mat{R}(\vect{q}_d) \vect{d}_* \vect{d}_*^\top  \mat{R}^\top(\vect{q}_d)\bs{\nu} \\ 
\dot{\vect{f}}_{\Delta,2} &= \dot{\mat{R}}(\vect{q}_d) \vect{d}_*  f  = \mat{R}(\vect{q}_d) [\bs{\omega}_d]_\times \vect{d}_*   f  \\ 
& = - \mat{R}(\vect{q}_d) [ \vect{d}_* ]_\times \left[\vect{d}_* \right]_\times \mat{R}^\top(\vect{q}_d)\bs{\nu} \\
\dot{\vect{f}}_{\Delta,3} &= - \dot{\vect{f}}_r =   k_{pp} \dot{\vect{e}}_p + k_{pd} \dot{\vect{e}}_v  \\
&=   k_{pp} {\vect{e}}_v + \frac{k_{pd}}{m} \left(-k_{pp}\vect{e}_p-k_{pd}\vect{e}_v+\vect{f}_\Delta \right) 
\end{align}
where we used the  selections of $\bs{\omega}_d, \dot{f}$ in~\eqref{eq:omega_d}, \eqref{eq:f_dot}, respectively, and $\vect{f}_r$ in~\eqref{eq:f_r}.
Employing~\eqref{eq:nu}, it follows that 
\begin{align}
\dot{\vect{f}}_{\Delta} &= \bs{\nu} -  \frac{k_{pd} k_{pp}}{m} \vect{e}_p
-\left(\frac{k_{pd}^2}{m}-k_{pp}\right) \vect{e}_v +  \frac{k_{pd} }{m}  \vect{f}_\Delta \\
&= -k_\Delta \vect{f}_\Delta,
\label{eq:f_Delta_dot2}
\end{align} 
%
It can be observed that the relation $\dot{\vect{f}}_\Delta\!=\!-k_\Delta \vect{f}_\Delta$  in~\eqref{eq:f_Delta_dot2} clearly establishes the exponential stability of ${\mathcal Z}_f$ near ${\mathcal Z}_0$ for the dynamics restricted to $ {\mathcal Z}_{a}$, using the Lyapunov function $V_\Delta:=\vect{f}_\Delta^\top \vect{f}_\Delta$.
\end{myproof}

As a final step, let us consider the set ${\mathcal Z}_0$ introduced in~\eqref{eq:A0} and restrict the attention to initial conditions in the set ${\mathcal Z}_f$. We can establish the next result.
\begin{lemma} \label{lem:A_0}
Set ${\mathcal Z}_0$ is asymptotically stable for the closed-loop dynamics, relative to initial conditions in ${\mathcal Z}_f$.
\end{lemma}
\begin{myproof}
Consider dynamics \eqref{eq:poserror_dyn1}-\eqref{eq:poserror_dyn2} for initial conditions in ${\mathcal Z}_f \subset  {\mathcal Z}_{a}$. Such dynamics corresponds to the situation of input $\vect{f}_r$ acting directly on the translational component of the plant \eqref{eq:motion3}, therefore exponential stability is easily established by using the Lyapunov function
\begin{align}
V_p & := \frac{1}{2} m \vect{e}_v^\top\vect{e}_v+\frac{1}{2} k_{pp}\vect{e}_p^\top\vect{e}_p,
\end{align}
for which 
it is easy to verify that along the dynamics restricted to ${\mathcal Z}_f$ we get
\begin{align}
\dot V_p &= m \vect{e}_v^\top \dot{\vect{e}}_v  + k_{pp}\vect{e}_p^\top \dot{\vect{e}}_p \\
&= \vect{e}_v^\top(-mg\vect{e}_3+\vect{f}_r) +k_{pp}\vect{e}_p^\top\vect{e}_v\\
&= \vect{e}_v^\top(-k_{pp}\vect{e}_p -k_{pd}\vect{e}_v) +k_{pp}\vect{e}_p^\top\vect{e}_v\\
&=-k_{pd} \Vert \vect{e}_v\Vert^2.
\end{align}
Applying the invariance principle, we obtain that the following set is asymptotically stable relative to ${\mathcal Z}_f$
\begin{align}
\label{eq:Aq}
{\mathcal Z}_q &:= 
\big\{ \vect{z} \in {\mathcal Z} \; | \; \vect{q}_\Delta = \vect{q}_I,  \bs{\omega}_\Delta= \vect{0},\vect{f}_\Delta=\vect{0},\nonumber\\
& \hspace*{4.5cm} \vect{e}_t=\vect{0}, \vect{q} \in {\mathbb S}^3 \big\}.
\end{align}
Now observe that in ${\mathcal Z}_q$, we have from \eqref{eq:nu} that $\bs{\nu}=\vect{0}$. Then from \eqref{eq:omega_d} it follows $\bs{\omega}_d=\vect{0}$ and,  since $\bs{\omega}_\Delta= \vect{0}$, also
$\bs{\omega} = \mat{R}^\top \! (\vect{q}) \bs{\omega}_d = \vect{0}$, meaning that the attitude $\vect{q}$ is constant in ${\mathcal Z}_q$. Using $\vect{f}_\Delta = \vect{0}$ and $\vect{q}_{\Delta} = \vect{q}_I$ (which implies $\mat{R}(\vect{q}) = \mat{R}(\vect{q}_d)$), we obtain from \eqref{eq:f_Delta},
$\mat{R}(\vect{q})\vect{d}_* f = mg \vect{e}_3$, which clearly implies $\vert f \vert= mg$. These derivations entail that ${\mathcal Z}_q = {\mathcal Z}_0$, thus completing the proof.
\end{myproof}

The stated lemmas establish a cascaded-like structure of the error dynamics composed of three hierarchically related subcomponents converging to suitable closed and forward invariant nested subsets of the space $\mathcal{Z}$ where the variable $\vect{z}$ in~\eqref{eq:xidef} evolves. These three closed subsets are ${\mathcal Z}_0 \subset {\mathcal Z}_f$, ${\mathcal Z}_f \subset {\mathcal Z}_a$ and ${\mathcal Z}_a \subset \mathcal{Z}$, where the smallest one, ${\mathcal Z}_0$, is also compact.
Such a hierarchical structure well matches the stability results established in \cite[Prop. 14]{Maggiore2013} whose conclusion, together with the results of Lemmas~\ref{lem:A_a}-\ref{lem:A_0} implies the following main result of our paper.
\begin{theorem}
Consider the closed-loop system 
in Figure~\ref{fig:controller}
between plant \eqref{eq:motion1}-\eqref{eq:motion4} and the controller presented in Section~\ref{sec:controller}. The compact set ${\mathcal Z}_0$ in~\eqref{eq:A0} is asymptotically stable for the corresponding dynamics.
\end{theorem}

\subsection{Extension}
\label{sec:ext}

The control goal of Problem~\ref{prob:hovering} can be extended with an additional requirement of restricted stabilization of a given \textit{reference orientation} $\vect{q}_r \in \mathbb{S}^3$ (where `restricted' refers to the fact that such an orientation should be tracked with a lower hierarchical priority as compared to the translational error stabilization).

For this extended goal, it is possible to modify the expression of $\bs{\omega}_d$ in order to exploit all the available degrees of freedom. Specifically, an additional term could be introduced in~\eqref{eq:omega_d} to asymptotically control the platform rotation around direction $\vect{d}_*$, with the aim of minimizing the mismatch between $\vect{q}_d$ and $\vect{q}_r$. To this end, we consider the following quantity in $\mathbb{S}^3$
\begin{align}
\label{eq:q_Delta_tilde}
{\vect{q}}_\Delta^\prime \!:=\! \vect{q}_r^{-1} \otimes \vect{q}_d \!=\! \colvec{
\eta_r \eta_d + \eps_r^\top \eps_d \\ -\eta_d \eps_r +\eta_r \eps_d - [\eps_r]_\times \eps_d 
} \!=\! \colvec{
{\eta}_\Delta^\prime \\ {\eps}_\Delta^\prime
}. \hspace{-0.1cm}
\end{align}
%
Then, the extended control goal can be achieved by replacing expression \eqref{eq:omega_d} by the following alternative form
\begin{align}
\bs{\omega}_d &= \frac{1}{f} \left[\vect{d}_* \right]_\times \mat{R}(\vect{q}_d)^\top \bs{\nu} + {\bs{\omega}}_d^\prime, \quad \text{with} \label{eq:omega_d_tilde}\\
{\bs{\omega}}_d^\prime &= -k_q \vect{d}_*\vect{d}_*^\top {\eps}_\Delta^\prime, \label{eq:omega_d_prime}
\end{align}
where $k_q \in \mathbb{R}^+$ is a proportional gain. The projection $ \vect{d}_*\vect{d}_*^\top$ in~\eqref{eq:omega_d_prime} is needed to ensure that the additional term does not influence the translational dynamics~\eqref{eq:motion3}, thereby encoding the hierarchical structure of the extended control goal. In other words, the orientation $\vect{q}_r$ is
obtained at the best maintaining the translational error of the platform equal to zero. 
Indeed, it is easy to verify that choice~\eqref{eq:omega_d_tilde} keeps  expression~\eqref{eq:f_Delta_dot2} of $\dot{\vect{f}}_\Delta$ unchanged.
On the other hand, it should be noted that expression \eqref{eq:omega_dd} will show an additional term, once the extended version of \eqref{eq:omega_d} is considered. 

The effectiveness of selection~\eqref{eq:omega_d_tilde} towards restricted tracking of
orientation $\vect{q}_r$ can be well established by using the Lyapunov function ${V}_\Delta^\prime  = 2{\eta}_\Delta^\prime$. Following the nested proof technique based on reduction theorems, it is enough to verify the negative semi-definiteness of the Lyapunov function derivative 
in the set $\mathcal{Z}_0$, where $\bs{\nu} = \vect{0}$ and $\bs{\omega}_d ={\bs{\omega}}_d'$. Then, using~\eqref{eq:q_d_dot}, \eqref{eq:q_Delta_tilde}, \eqref{eq:omega_d_tilde}, it follows that
\begin{align}
\dot{V}_\Delta^\prime &= 2 \dot{{\eta}}_\Delta^\prime  = ({\eps}_\Delta^\prime)^\top  \bs{\omega}_d \\
&=  -k_q ({\eps}_\Delta^\prime)^\top  \vect{d}_* \vect{d}_*^\top {\eps}_\Delta^\prime   = -k_q \Vert \vect{d}_*^\top {\eps}_\Delta^\prime\Vert^2 .
\end{align} 
Recalling that in set $\mathcal{Z}_0$ it holds that $\mat{R}(\vect{q}_d)\vect{d}_* = \vect{e}_3$, the above analysis reveals that asymptotically
one obtains $ \vect{d}_*^\top {\eps}_\Delta^\prime =0$, which seems to suggest that there is some control achievement (within the restricted goal) in the direction orthogonal to $\vect{d}_*$ (resembling a steady-state yaw direction).

\section{Simulation Results}
\label{sec:simulation}

\begin{figure}[t]
\centering
\includegraphics[scale=0.5]{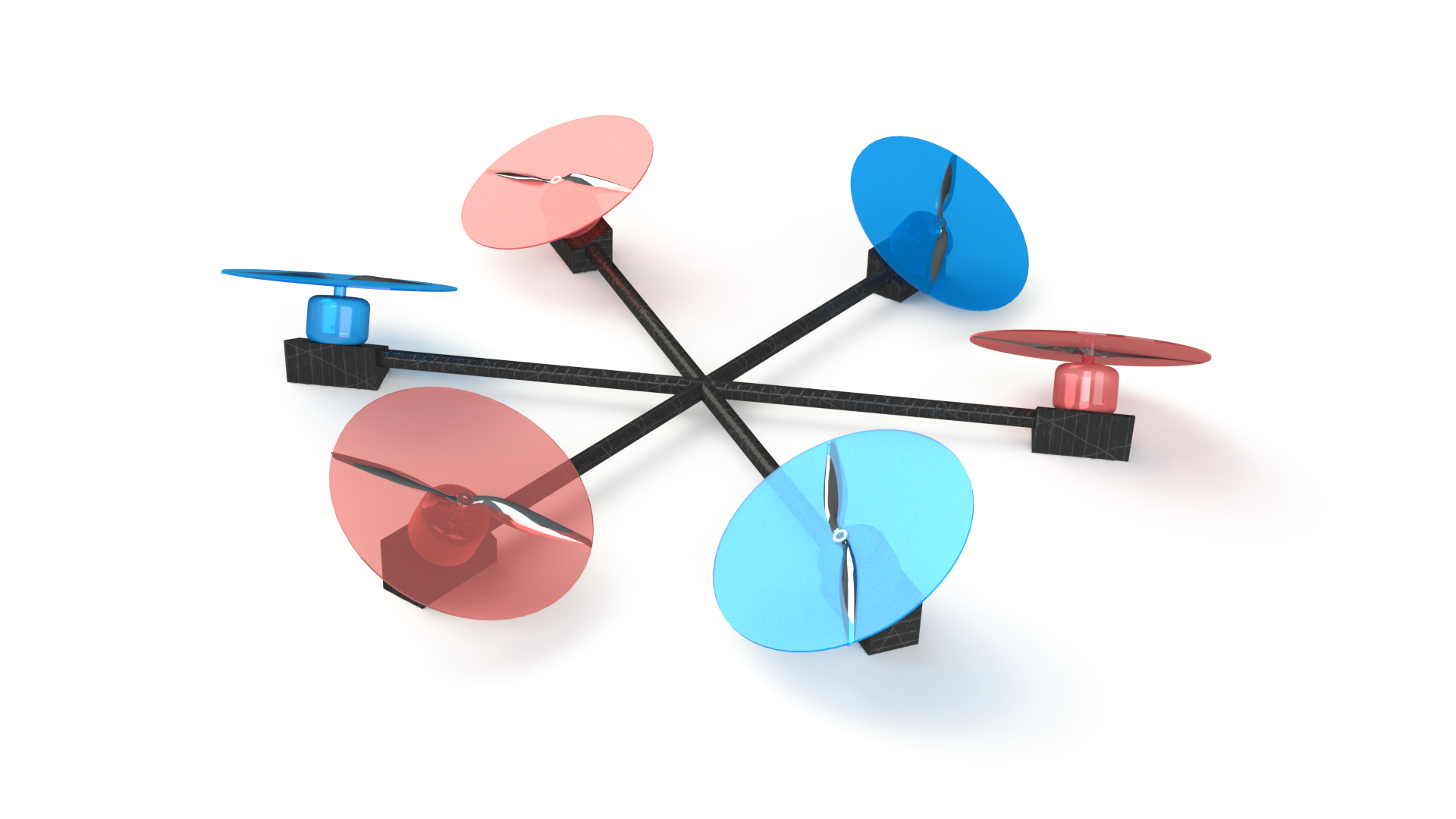} 
\caption{Star-shaped hexarotor with tilted propellers described in Section~\ref{sec:simulation} - red/blue discs correspond to CW/CCW rotors.}
\label{fig:hexa}
\end{figure}

The effectiveness of the proposed controller for solving Problem~\ref{prob:hovering} is here validated by numerical simulations on a specific instantiation of hexarotor introduced in~\cite{Rajappa2015} characterized by $n=6$ tilted propellers having the same geometric and aerodynamics features (i.e., $c_{f_i}= c_f$ and $c_{\tau_i} = c_\tau$, $i=1 \ldots 6$). This is depicted in Figure~\ref{fig:hexa}.

To exhaustively describe the platform, we consider the frame $\mathscr{F}_{P_i} = \{ O_{P_i}, (\vect{x}_{P_i}, \vect{y}_{P_i}, \vect{z}_{P_i})\}$ for each rotor  $i = 1 \ldots 6$. The origin $O_{P_i}$ coincides with the CoM of the $i$-th motor-propeller combination,  $\vect{x}_{P_i}$ and $\vect{y}_{P_i}$ identify its spinning plane, while $\vect{z}_{P_i}$ coincides with its spinning axis. 
As shown in Figure~\ref{fig:hexa}, $O_{P_1} \ldots O_{P_6}$ lie on the same plane where they are equally spaced along a circle, namely we account for  a  \textit{star-shaped} hexarotor. Formally, for $i = 1 \ldots n$, the position $\vect{p}_i \in \mathbb{R}^3$ of $O_{P_i}$ in $\mathscr{F}_{B}$ is set as
\begin{align}
\vect{p}_i = \vect{q}(\gamma_i,\vect{e}_3) \otimes   \colvec{
0 & \ell & 0 & 0 
}^\top \otimes \vect{q}(\gamma_i,\vect{e}_3)^{-1}
\end{align}
where $\vect{q}(\gamma_i,\vect{e}_3) \in \mathbb{S}^3
$ is the unit quaternion associated to the rotation by $\gamma_i=(i-1)\pi/3$ about $\vect{e}_3$ according to the axis-angle representation given in Section~\ref{sec:preliminaries}, and $\ell>0$ is the distance between $O_{P_i}$ and $O_B$.  Moreover, we  assume that the orientation of each $\mathscr{F}_{P_i}$ w.r.t. $\mathscr{F}_B$ can be represented by the unit quaternion $\vect{q}_i \in \mathbb{S}^3$ such that
\begin{align}
\vect{q}_i = \vect{q}(\gamma_i,\vect{e}_3) \otimes   \vect{q}(\beta_i,\vect{e}_2) \otimes \vect{q}(\alpha_i,\vect{e}_1) 
\end{align}
where $\vect{q}(\beta_i,\vect{e}_2), \vect{q}(\alpha_i,\vect{e}_1) \!\!\in \!\!\mathbb{S}^3$ agree with the axis-angle representation and the tilt angles $\alpha_i, \beta_i  \!\!\in \!\! (-\pi,\pi]$ uniquely define the direction of $\vect{z}_{P_i}$ in $\mathscr{F}_B$. Indeed, the frame $\mathscr{F}_{P_i}$ is obtained from $\mathscr{F}_{B}$ by first rotating by $\alpha_i$ about $\vect{x}_{B}$ and then by $\beta_i$ around $\vect{y}_{B}'$. In particular, these angles are chosen so that $\alpha_{i}=-\alpha_{i+1}$ and $\alpha_{i}\neq \alpha_{j}$ for $i,j=1,3,5$, while $\beta_i = \beta $ for $i=1\ldots 6$. 

The choice of this complex and rather anomalous configuration is motivated by the fact that it can realize the static hovering condition and satisfies the Assumption~\ref{as:standing}, but the matrix $\mat{K}$ in~\eqref{eq:u} is not trivially the identity matrix. Nevertheless, $\mat{K}$ can be chosen as the product between an orthogonal basis of the null space of $\mat{F}$ and its transpose (i.e., $\mat{K} = \bar{\mat{F}} (\bar{\mat{F}})^\top$ as in the proof of Lemma~\ref{lemma:equivalence}). 

The performed simulation exploits the dynamic model~\eqref{eq:motion1}-\eqref{eq:motion4} extended by several real-world effects.
\begin{itemize}
\item The position and orientation feedback and their derivatives are affected by time delay $t_f=\SI{0.012}\second$ and Gaussian noise corrupts the measurements according to Table~\ref{tab:sensorNoise}. The actual position and orientation are fed back with a lower sampling frequency of $\SI{100}{\hertz}$ while the controller runs at $\SI{500}{\hertz}$. These properties are reflecting a typical motion capture system and an inertial measurement unit (IMU).\\
\item The electronic speed controller (ESC) driving the motors is simply modeled by quantizing the desired input $\vect{u}$ resembling a ${10}$\,bit discretization in the feasible motor speed resulting in a step size of $\approx\SI{0.12}{\hertz}$. Additionally, the motor-propeller combination is modeled as a first order transfer function $\left(G(s)=(1+0.005s)^{-1}\right)$. The resulting signal is corrupted by a rotational velocity dependent Gaussian noise (see Table~\ref{tab:sensorNoise}). This combination reproduces quite accurately the dynamic behavior of a common ESC motor-propeller combination, i.e., BL-Ctrl-2.0, by MikroKopter, Robbe ROXXY 2827-35 and a $10$\,inch rotor blade~\cite{Franchi2017}.
\end{itemize}

\begin{table}[t]
\begin{center}
\caption{Standard deviation of the modeled sensor noise added to the corresponding measurements.}
\resizebox{\columnwidth}{!}{%
  \begin{tabular}{ | c | c | c | c |}
    \hline
    $\vect{p}$ & ${\vect{v}}$ & $\vect{q}$ & $\bs{\omega}$  \\ \hline \hline 
    $\SI{6.4e-04}{\meter}$ & $\SI{1.4e-03}{\meter/\second}$ & $\SI{1.2e-03}{}$ &  $\SI{2.7e-03}{\radian/\second}$ \\ \hline
  \end{tabular}}
\label{tab:sensorNoise}
\end{center}
\end{table}


\begin{figure}[t!]
\centering
\subfloat{\includegraphics[width=0.98\columnwidth]{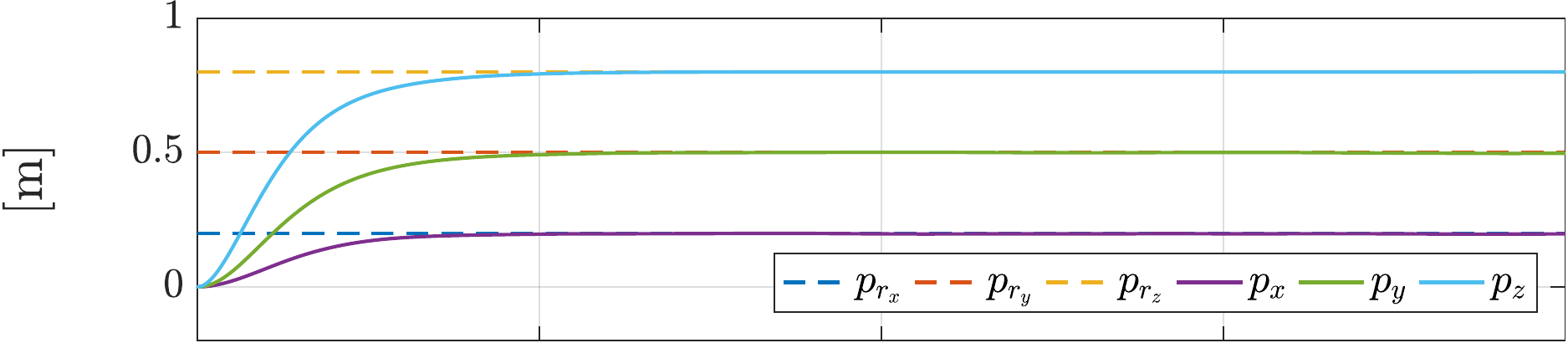}}\\
\subfloat{\includegraphics[width=0.98\columnwidth]{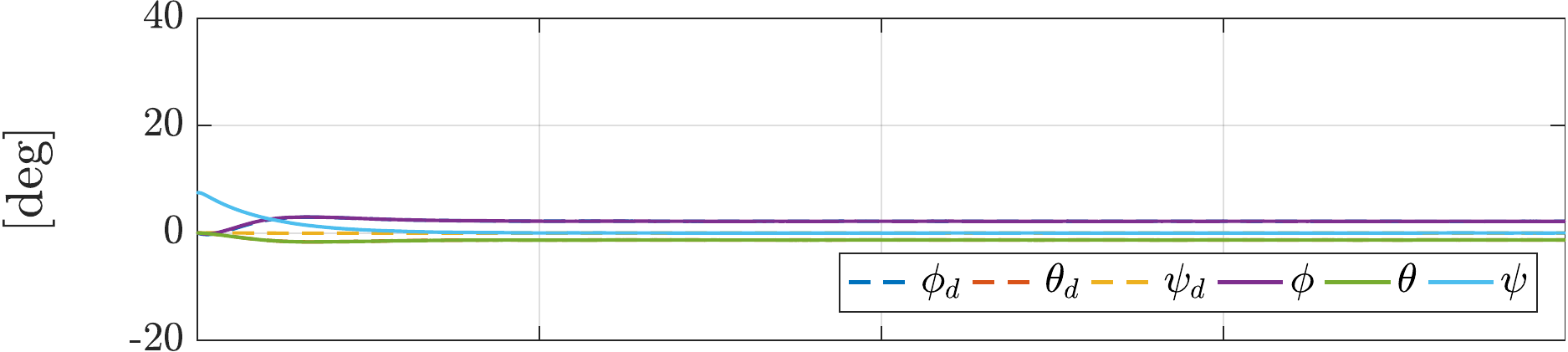}}\\
\subfloat{\includegraphics[width=0.98\columnwidth]{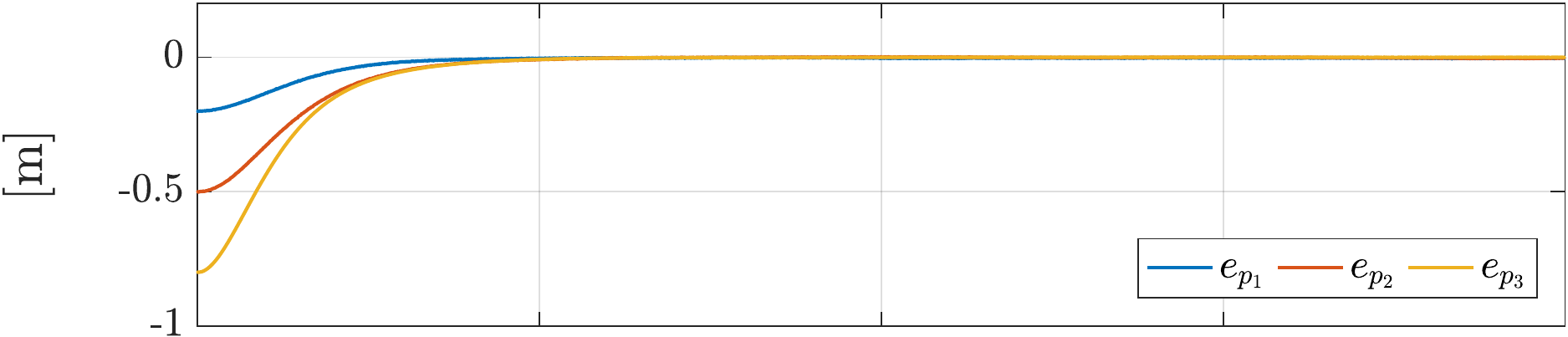}} \\
\subfloat{\includegraphics[width=0.98\columnwidth]{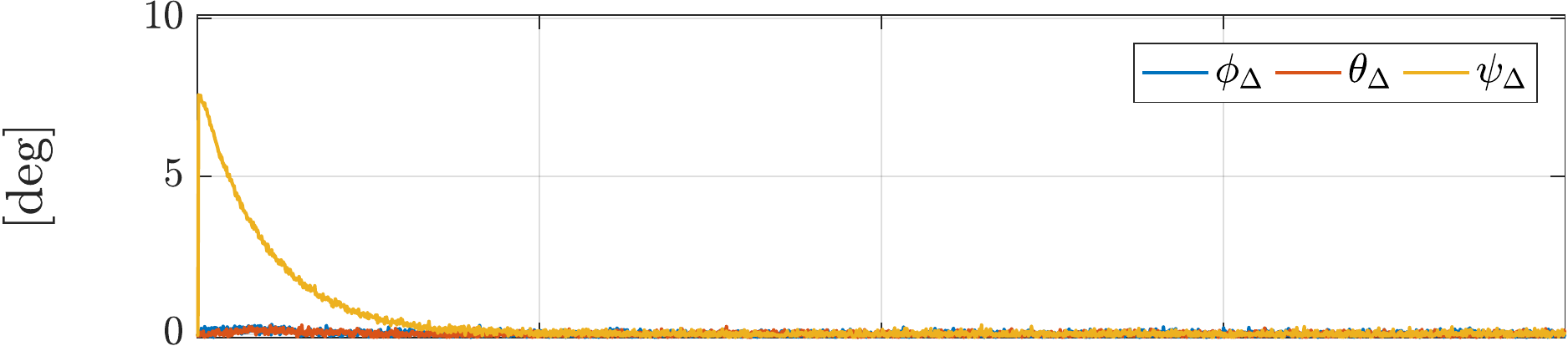}}\\
\subfloat{\includegraphics[width=0.995\columnwidth]{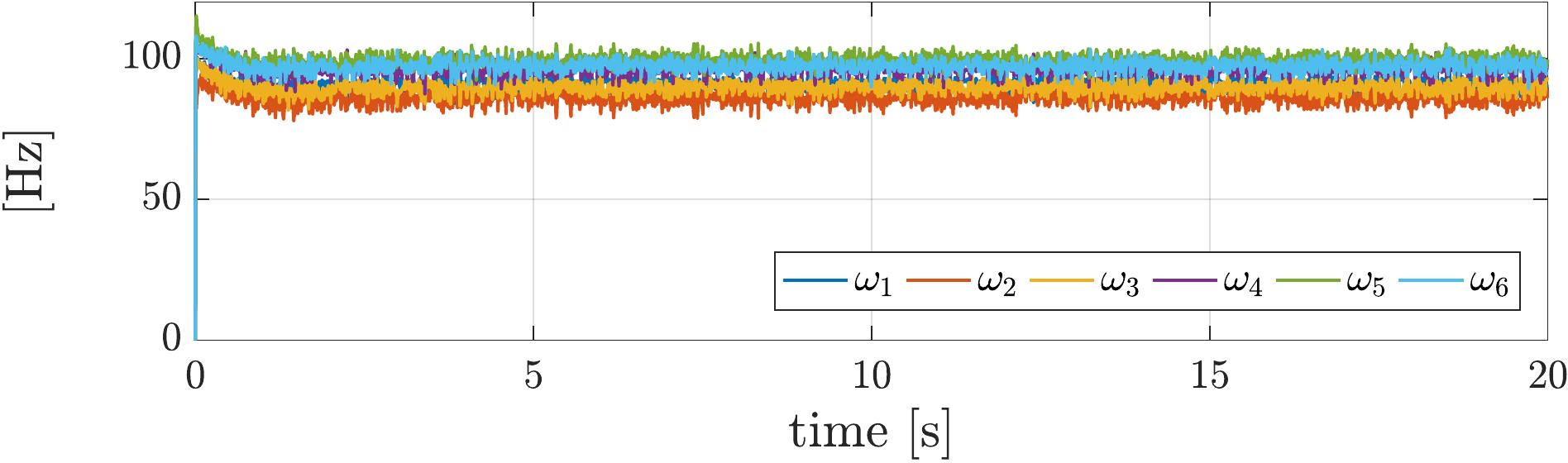}}
\caption{Hover control of the hexarotor in real conditions.}
\label{fig:p_r_only}
\end{figure}

\begin{figure}[t!]
\centering
\subfloat{\includegraphics[width=0.98\columnwidth]{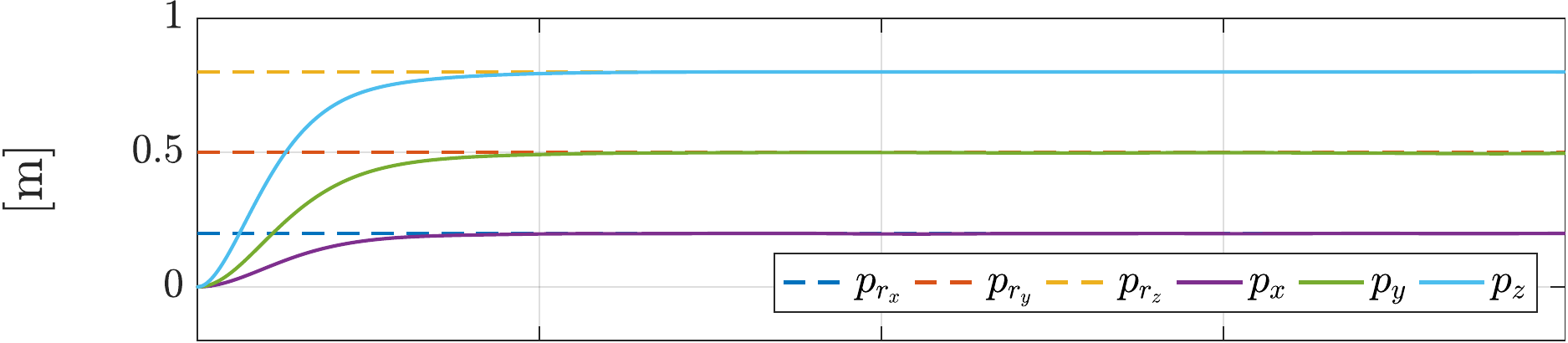}}\\
\subfloat{\includegraphics[width=0.98\columnwidth]{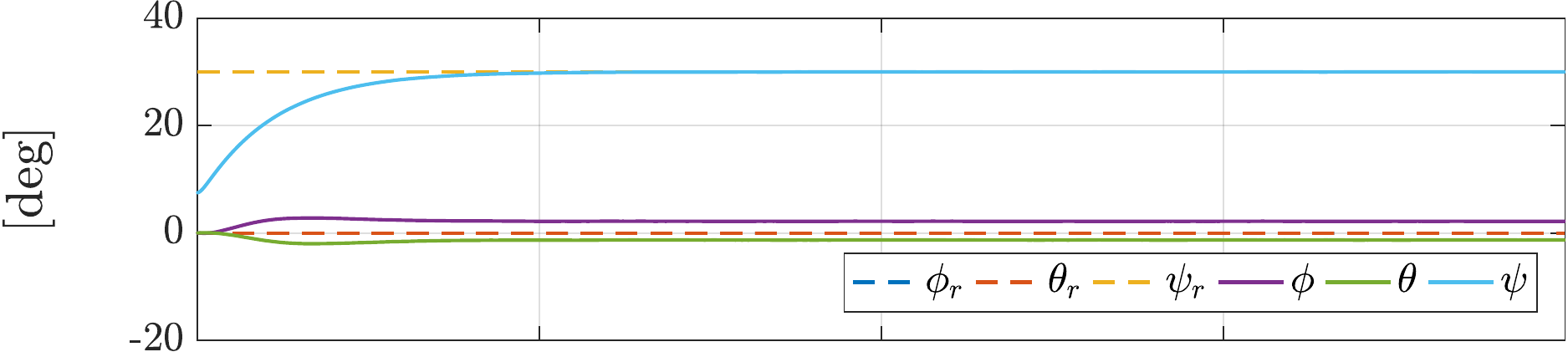}}\\
\subfloat{\includegraphics[width=0.98\columnwidth]{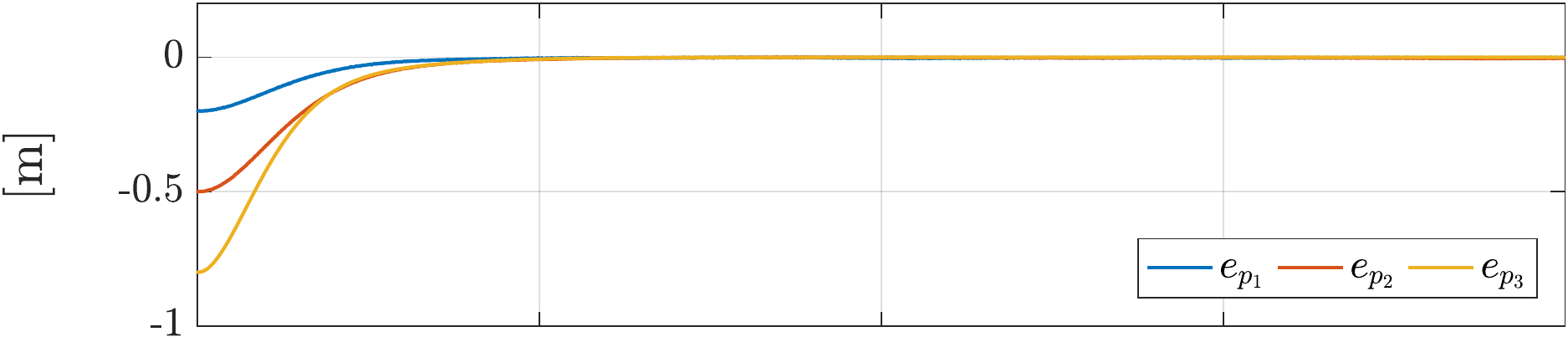}} \\
\subfloat{\includegraphics[width=0.98\columnwidth]{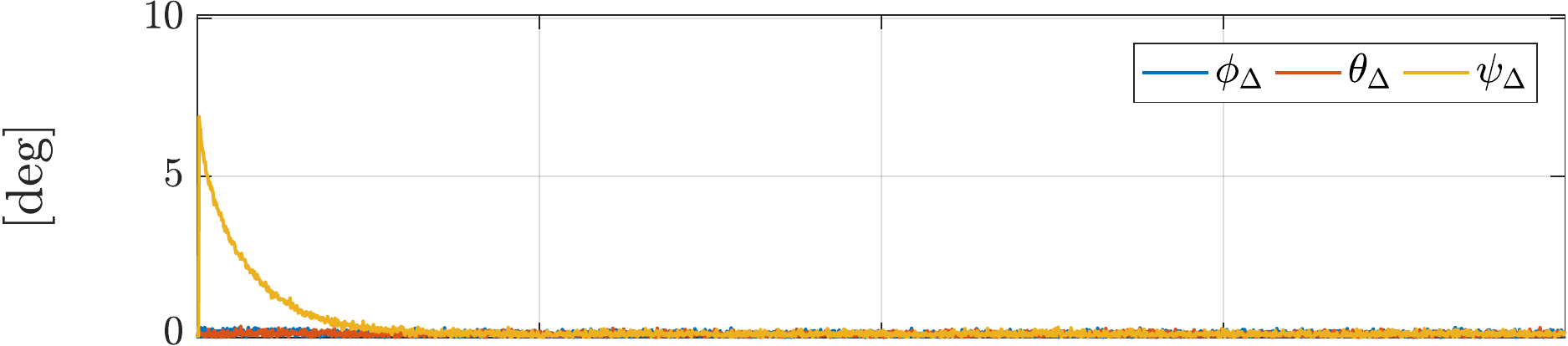}}\\
\subfloat{\includegraphics[width=0.995\columnwidth]{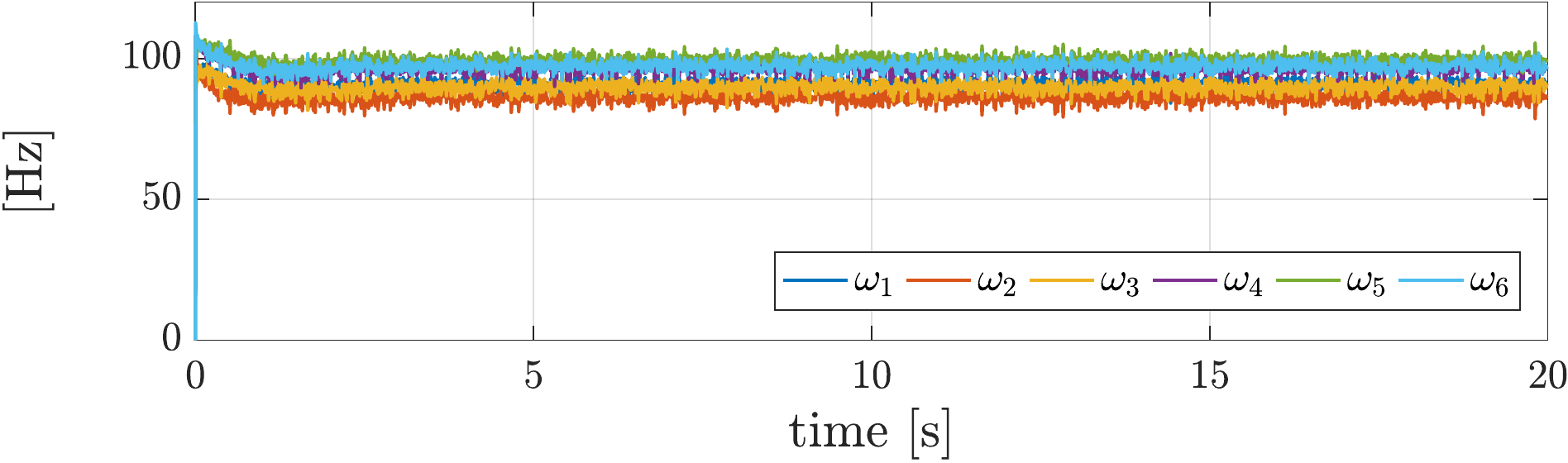}}
\caption{Hover control of the hexarotor in real conditions providing a constant reference orientation.}
\label{fig:p_r_q_r}
\end{figure}

The control goal is firstly to steer the described vehicle to a locally stable equilibrium position $\vect{p}_r\in \mathbb{R}^3$ without imposing a reference orientation.
The simulation results are depicted in Figure~\ref{fig:p_r_only}. The first and second plot report the position and orientation of the hexarotor, respectively. The roll-pitch-yaw angles $(\phi,\theta,\psi)$ are used to represent the attitude to give a better insight of the vehicle behavior, however, the internal computations are all done with unit quaternions. The hexarotor smoothly achieves the reference position in roughly $\SI{5}\second$. After this transient, the position error $\vect{e}_p$ (third plot) converges to zero. This behavior is expected in light of the robustness results of asymptotic stability of compact attractors, established in~\cite[Chap.~7]{Goebel2012}. Similarly, the orientation $\vect{q}$ of the vehicle converges to the desired one $\vect{q}_d$  with a comparable transient time scale. This is clearly visible in the fourth plot that reports the trend of the roll-pitch-yaw angles associated to $\vect{q}_\Delta$. Note that $\vect{q}_d = \vect{q}_I$.  
The last plot in Figure~\ref{fig:p_r_only} shows the control inputs commanded to the propellers: at the steady-state, all the spinning rates are included in $[80,110]\si{\hertz}$, which represents a feasible range of values from a practical point of view.  

Figure~\ref{fig:p_r_q_r} illustrates the performance of the controller when a constant given orientation is required according to Section~\ref{sec:ext}. The error trends and the commanded spinning rates are comparable to the previous case, while the second plot shows that the hexarotor rotates according to the given $\vect{q}_r$, although a very small bias ($\approx \SI{2}\degree$) is observable in the roll and pitch components. However, the fourth plot ensures that these at least converge toward the desired values: the roll-pitch-yaw angles related to $\vect{q}_\Delta$ converges toward zero ensuring that the current orientation $\vect{q}$ approximates the desired one $\vect{q}_d$, which results to be slightly different from the required $\vect{q}_r$.  

\section{Conclusions}\label{sec:conclusions}

We addressed the hovering control task for a generic class of multi-rotor vehicles whose propellers are arbitrary in number and spinning axis mutual orientation. Adopting the quaternion attitude representation, we designed a state feedback non-linear controller to stabilize a UAV in a reference position with an arbitrary but constant orientation. The proposed solution relies on some non-restrictive assumptions on the control input matrices $\mat{F}$ and $\mat{M}$ that ensure the existence of a preferential direction in the feasible force space, along which the control force and the control moment are decoupled.  Stability and asymptotic convergence of the tracking error has been rigorously proven through a cascaded-like proof exploiting nested sets and reduction theorems. The theoretical findings are confirmed by the numerical simulation results, supporting the test of the control scheme  on a real platform in the near future.

\appendix

\section{Proof of the identity $\dot{\bs{\omega}}_{{d}}=\bs{\omega}_{{dd}}$}
\label{sec:appendix}

The identity $\dot{\bs{\omega}}_d = \bs{\omega}_{dd}$ stated in Sec~\ref{sec:controller} is justified by in the following  where we exploit also the relation 
$\left[ [\eps_1]_\times \eps_2 \right]_\times = [\eps_1]_\times [\eps_2]_\times - [\eps_2]_\times [\eps_1]_\times  = \eps_2 \eps_1^\top -\eps_1 \eps_2^\top.$

The derivative of $\bs{\omega}_d$ in~\eqref{eq:omega_d} results from the sum of three components, namely $\dot{\bs{\omega}}_d = \dot{\bs{\omega}}_{d,1} + \dot{\bs{\omega}}_{d,2} + \dot{\bs{\omega}}_{d,3}$
with
\begin{align}
\dot{\bs{\omega}}_{d,1} &= - \frac{1}{ f^2}  [\vect{d}_*]_\times \mat{R}^\top_d \bs{\nu} \dot{f} \\
&\stackrel{\eqref{eq:f_dot}}{=}  - \frac{1}{ f^2}  [\vect{d}_*]_\times \mat{R}^\top_d \bs{\nu} \vect{d}_*^\top \mat{R}_d^\top  \bs{\nu}  \\
&= - \frac{\left(\vect{d}_*^\top \mat{R}_d^\top  \bs{\nu} \right) }{ f^2} [\vect{d}_*]_\times \mat{R}^\top_d \bs{\nu} \\ 
\dot{\bs{\omega}}_{d,2} &= \frac{1}{f}   [\vect{d}_*]_\times \dot{\mat{R}}^\top_d \bs{\nu} \\ 
&=  - \frac{1}{f}  [\vect{d}_*]_\times [\bs{\omega}_d]_\times {\mat{R}}^\top_d \bs{\nu}  \\
& \stackrel{\eqref{eq:omega_d}}{=}  -\frac{1}{ f^2}  [\vect{d}_*]_\times \left[ \left[ \vect{d}_* \right]_\times \mat{R}^\top_d \bs{\nu}\right]_\times {\mat{R}}^\top_d \bs{\nu}\\ \displaybreak
%
& 
=- \frac{1}{ f^2}  [\vect{d}_*]_\times  \mat{R}^\top_d \bs{\nu} \vect{d}_*^\top  {\mat{R}}^\top_d \bs{\nu} \\ 
&= - \frac{\left(\vect{d}_*^\top \mat{R}_d^\top  \bs{\nu} \right) }{ f^2} [\vect{d}_*]_\times \mat{R}^\top_d \bs{\nu}
\end{align}
where $\mat{R}^\top_d$ stands for $\mat{R}^\top(\vect{q}_d)$. Thus, we get 
\begin{align}
\dot{\bs{\omega}}_{d,1}+\dot{\bs{\omega}}_{d,2} &= - \frac{2}{ f^2}\left(\vect{d}_*^\top \mat{R}_d^\top  \bs{\nu} \right)  [\vect{d}_*]_\times \mat{R}^\top_d \bs{\nu}, \\
&= - \frac{1}{f}\kappa(\vect{e}_p,\vect{e}_v,\vect{f}_\Delta) [\vect{d}_*]_\times \mat{R}^\top_d \bs{\nu}, \label{eq:sum_1}
\end{align}
by introducing the gain $\kappa(\vect{e}_p,\vect{e}_v,\vect{f}_\Delta) \in \mathbb{R}$ that, exploiting~\eqref{eq:nu}, results as in~\eqref{eq:kappa}.
The derivation of $\dot{\bs{\omega}}_{d,3}$ is instead reported  in~\eqref{eq:omega_d3}-\eqref{eq:omega_d3_fin} where $\mat{R}_d = \mat{R}(\vect{q}_d)$ and $\mat{R} = \mat{R}(\vect{q})$ to simplify the notation.
%
%

Using~\eqref{eq:sum_1}~and~\eqref{eq:omega_d3_fin}, and setting $k_1$, $k_2(\vect{e}_p,\vect{e}_v,\vect{f}_\Delta)$, $k_3(\vect{e}_p,\vect{e}_v,\vect{f}_\Delta)$ and $k_4(\vect{e}_p,\vect{e}_v,\vect{f}_\Delta)$ as in~\eqref{eq:k_1}-\eqref{eq:k_4}, it is trivial to verify that it results $\dot{\bs{\omega}}_d = \bs{\omega}_{dd}$.
\begin{figure*}[t]
\begin{small}
\begin{align}
\label{eq:omega_d3}
\dot{\bs{\omega}}_{d,3} &= \frac{1}{\xi f}  [\vect{d}_*]_\times   \mat{R}^\top_d \dot{\bs{\nu}} \\
& \stackrel{\eqref{eq:nu}}{=} \frac{1}{\xi f}  [\vect{d}_*]_\times  \mat{R}^\top(\vect{q}_d) \left( \frac{k_{pd} k_{pp}}{m}\dot{\vect{e}}_p+\left(  \frac{k_{pd}^2}{m}-k_{pp}\right) \dot{\vect{e}}_v 
 - \left(\frac{k_{pd} }{m}+k_\Delta \right) \dot{\vect{f}}_\Delta \right)  \\ 
& \stackrel{\eqref{eq:poserror_dyn2}}{=}    \frac{1}{\xi f}  [\vect{d}_*]_\times  \mat{R}^\top_d \left( \frac{k_{pd} k_{pp}}{m}{\vect{e}}_v  - \left(\frac{k_{pd} }{m}+k_\Delta \right) \dot{\vect{f}}_\Delta+ \left(  \frac{k_{pd}^2}{m^2}-\frac{k_{pp}}{m}\right) \left(-mg\vect{e}_3 + (\mat{R} - \mat{R}_d \vect{d}_* \xi f + \vect{f}_r + \vect{f}_\Delta\right)
 \right)   \\ 
 & =   \frac{1}{\xi f}  [\vect{d}_*]_\times  \mat{R}^\top_d \left( \frac{k_{pd} k_{pp}}{m}{\vect{e}}_v  - \left(\frac{k_{pd} }{m}+k_\Delta \right) \dot{\vect{f}}_\Delta+ \left(  \frac{k_{pd}^2}{m^2}-\frac{k_{pp}}{m}\right) \left(-mg\vect{e}_3 + \mat{R}\vect{d}_* \xi f + \vect{f}_r + \vect{f}_\Delta\right)
 \right)   \\ 
 & \stackrel{\eqref{eq:f_r}}{=}    \frac{1}{\xi f}  [\vect{d}_*]_\times  \mat{R}^\top_d \left( \frac{k_{pd} k_{pp}}{m}{\vect{e}}_v  - \left(\frac{k_{pd} }{m}+k_\Delta \right) \dot{\vect{f}}_\Delta +\left(  \frac{k_{pd}^2}{m^2}-\frac{k_{pp}}{m}\right) \left( \mat{R}\vect{d}_* \xi f - k_{pp} \vect{e}_p - k_{pd}\vect{e}_v + \vect{f}_\Delta\right) 
 \right)   \\ 
 & \stackrel{\eqref{eq:f_Delta_dot2}}{=}    \frac{1}{\xi f}  [\vect{d}_*]_\times  \mat{R}^\top_d \left( \frac{k_{pd} k_{pp}}{m}{\vect{e}}_v  + \left(\frac{k_{pd} }{m}+k_\Delta \right) k_\Delta {\vect{f}}_\Delta +\left(  \frac{k_{pd}^2}{m^2}-\frac{k_{pp}}{m}\right) \left( \mat{R}\vect{d}_* \xi f - k_{pp} \vect{e}_p - k_{pd}\vect{e}_v + \vect{f}_\Delta\right) 
 \right)   \label{eq:omega_d3_fin}
\end{align}
\end{small}
\vspace{-0.3cm}
\hrule
\vspace{0.0cm}
\end{figure*}

\bibliographystyle{plain}
\bibliography{./bibCustom2}

\end{document}